\documentclass[sigconf]{acmart}

\AtBeginDocument{%
  \providecommand\BibTeX{{%
    \normalfont B\kern-0.5em{\scshape i\kern-0.25em b}\kern-0.8em\TeX}}}

\usepackage{multirow}
\usepackage{array}
\usepackage{tabularx,booktabs}
\newcolumntype{Y}{>{\centering\arraybackslash}X}
\usepackage{enumitem}
\newcommand{\proposed}{\textsf{TERACON}}
\usepackage{xcolor}
\usepackage{pifont}
\newcommand{\cmark}{\textcolor{blue}{\ding{51}}}%
\newcommand{\xmark}{\textcolor{red}{\ding{55}}}%
\setcopyright{acmcopyright}
\copyrightyear{2023}
\acmYear{2023}
\setcopyright{acmlicensed}
\copyrightyear{2023}
\acmYear{2023}
\setcopyright{acmlicensed}\acmConference[KDD '23]{Proceedings of the 29th ACM SIGKDD Conference on Knowledge Discovery and Data Mining}{August 6--10, 2023}{Long Beach, CA, USA}
\acmBooktitle{Proceedings of the 29th ACM SIGKDD Conference on Knowledge Discovery and Data Mining (KDD '23), August 6--10, 2023, Long Beach, CA, USA}
\acmPrice{15.00}
\acmISBN{979-8-4007-0103-0/23/08}
\acmDOI{10.1145/3580305.3599516}

\begin{document}

\title{Task Relation-aware Continual User Representation Learning}

\author{Sein Kim}
\authornote{Work done while the authors were interning at NAVER Corporation.}
\affiliation{%
  \institution{KAIST}
  \country{Republic of Korea}
}
\email{rlatpdlsgns@kaist.ac.kr}

\author{Namkyeong Lee} 
\authornotemark[1]
\affiliation{
\institution{KAIST}
  \country{Republic of Korea}
}
\email{namkyeong96@kaist.ac.kr}

\author{Donghyun Kim}
\affiliation{
\institution{NAVER Corporation}
  \country{Republic of Korea}
}
\email{amandus.kim@navercorp.com}

\author{Minchul Yang}
\affiliation{
\institution{NAVER Corporation}
  \country{Republic of Korea}
}
\email{minchul.yang@navercorp.com}

\author{Chanyoung Park}
\authornote{Corresponding author.}
\affiliation{
\institution{KAIST}
  \country{Republic of Korea}
}
\email{cy.park@kaist.ac.kr}

\renewcommand{\shortauthors}{Sein Kim, Namkyeong Lee, Donghyun Kim, Minchul Yang, \& Chanyoung Park}

\begin{abstract}
      User modeling, which learns to represent users into a low-dimensional representation space based on their past behaviors, got a surge of interest from the industry for providing personalized services to users.
      Previous efforts in user modeling mainly focus on learning a task-specific user representation that is designed for a single task. However, since learning task-specific user representations for every task is infeasible, recent studies introduce
      the concept of universal user representation, which is a more generalized representation of a user that is relevant to a variety of tasks. 
      Despite their effectiveness, existing approaches for learning universal user representations are impractical in real-world applications due to the data requirement, catastrophic forgetting and the limited learning capability for continually added tasks.
      In this paper, we propose a novel continual user representation learning method, called \proposed, whose learning capability is not limited as the number of learned tasks increases while capturing the relationship between the tasks.
      The main idea is to introduce an embedding for each task, i.e., \textit{task embedding}, which is utilized to generate task-specific soft masks that not only allow the entire model parameters to be updated until the end of training sequence, but also facilitate the relationship between the tasks to be captured.
      Moreover, we introduce a novel knowledge retention module with pseudo-labeling strategy that successfully alleviates the long-standing problem of continual learning, i.e., catastrophic forgetting.
      Extensive experiments on public and proprietary real-world datasets demonstrate the superiority and practicality of \proposed.
      Our code is available at \url{https://github.com/Sein-Kim/TERACON}.
\end{abstract}





\maketitle

\section{Introduction}
To identify relevant information from the overloaded data, machine learning (ML) has become a crucial tool for social media and e-commerce platforms, boosting not only user experience but also business revenue \cite{10.1145/2959100.2959190}.
Among the various ML techniques, user modeling (UM) got a surge of interest from the industry due to its ability in discovering user's latent interests, thereby enabling personalized services for a variety of users \cite{10.1145/2806416.2806527,5197422,hidasi2015session,kang2018self,yuan2019simple, kim2023melt}.
The key idea in UM is to represent users into a low-dimensional representation space based on their past behaviors, which inherently contain rich and diverse users' interests.


Despite the significant progress in UM, most of the previous efforts have been concentrated on learning task-specific user representations, which are specifically designed for a single task (e.g., click-through rate prediction, user profiling) \cite{guo2017deepfm,yuan2019simple,zhou2019deep}, limiting its applicability to real-world applications.
For example, companies like Amazon and Alibaba offer various services to users in a single platform, all of which require service-specific UM to enhance user satisfaction and improve business revenue \cite{sun2021interest, sheng2021one,chui2018notes}.
However, learning service-specific user representations for every service is impractical in reality due to expensive memory/financial costs.
To overcome the limitations of task-specific user representations, 
the concept of a universal user representation has recently emerged as a solution. 
The key idea is to learn a generalized representation of a user that better captures the user's underlying characteristics, which are relevant for not only single task but also across a variety of tasks \cite{yuan2020parameter, yuan2021one, Ni2018PerceiveYU, gu2021exploiting}.

Existing studies mainly adopt multi-task learning (MTL) \cite{crawshaw2020multi,ruder2017overview} to learn universal user representations.
MTL aims to learn a generalized representation of users by simultaneously training a single model for each user on multiple tasks~\cite{ni2018perceive,10.1145/3219819.3220007}.
However, MTL has inherent limitations as it requires all tasks and their associated data to be available in advance, which rarely holds in real-world situations. In other words, when introducing a new service, MTL-models need to be retrained on both the new data and all the data from previous tasks.
Meanwhile, Transfer learning (TL) offers an alternative approach to address this issue.
More precisely, given a pair of tasks, i.e., source task with abundant data and target task with insufficient data, TL aims to improve the performance on the target task by transferring the knowledge obtained from the source task, which is different but positively related to the target task \cite{yuan2020parameter,10.5555/2891460.2891628,yang2021autoft}.
Specifically, PeterRec \cite{yuan2020parameter} boosts the model performance on various target tasks by pre-training the model on the user-item interaction in a source task in a self-supervised manner.
However, TL is only applicable when a pair of tasks is given, whereas online platforms usually contain multiple target tasks. 
For this reason, even if multiple target tasks are positively related, i.e., offer useful information to each other, knowledge transfer can only happen between a source-target pair.
Moreover, the model easily suffers from catastrophic forgetting, that is, the model performance on the source task severely deteriorates after training the model on the target task \cite{mccloskey1989catastrophic,ratcliff1990connectionist}.

Recently, continual learning (CL) has shown great success in learning a single model on a stream of multiple tasks while retaining the knowledge from the tasks observed in the past~\cite{de2021continual,lesort2020continual,biesialska2020continual}, which alleviates the shortcomings of MTL and TL mentioned above. More precisely, CL is similar to MTL in that the model learns from multiple tasks, but CL does not require the simultaneous availability of training data across multiple target tasks.
{Besides, CL is similar to TL in that the model can transfer knowledge from previous tasks to new tasks. However, CL also retains the knowledge from previous tasks, and some CL models capture the relationship between multiple target tasks~\cite{ke2020continual,lee2021continual}.}
CONURE \cite{yuan2021one} is the first work to adopt the CL framework to recommender systems, and it learns a universal user representation based on a series of multiple tasks. 
The main idea is to train the model on a task using only a portion of the model parameters, which are then fixed when being trained on the next task where another portion of the remaining model parameters is used.
This enables the model to retain the knowledge obtained in the past, i.e., avoid catastrophic forgetting, and such an approach is called parameter isolation.

Despite its effectiveness, CONURE fails to learn from the continuously incoming sequence of tasks due to the inherent limitation of the parameter isolation approach, which restricts the modification of model parameters that are used in previous tasks.
While this approach prevents catastrophic forgetting, it limits the model's capability to learn subsequent tasks due to the gradual reduction of available learnable parameters as more tasks are added.
Another drawback of CONURE is that it does not consider the relationship between tasks.
For example, the knowledge obtained from predicting a user's age could enhance item purchase prediction, since certain items may be popular among users of a specific age, whereas the age prediction task would not be helpful for predicting the gender of a user.
Hence, capturing the relationship between tasks encourages a positive transfer between positively related tasks, while preventing a negative transfer between negatively related tasks \cite{zamir2018taskonomy, ke2020continual}, and thus it is crucial for an effective training of a new task.

To this end, we propose a novel \textsf{T}ask \textsf{E}mbedding-guided \textsf{R}elation-\textsf{A}ware \textsf{CON}tinual user representation learner, named \proposed, whose learning capability is not limited regardless of the number of continually added tasks, while capturing the relationship between tasks.
The main idea is to introduce an embedding for each task, i.e., \textit{task embedding}, which is utilized to generate a task-specific soft mask that not only allows the entire model parameters to be updated until the end of the training sequence thereby retaining the learning capability of the model, but also facilitates the relationship between the tasks to be captured. Moreover, to prevent catastrophic forgetting, we propose a knowledge retention module that  transfers the knowledge of the current model regarding previous tasks to help train the current model itself. Lastly, we propose a sampling strategy to make the training more efficient. 
Our extensive experiments on two real-world datasets and a proprietary dataset demonstrate that~\proposed~outperforms the state-of-the-art continual user representation learning methods. 

Our contributions are summarized as follows:
\begin{itemize}[leftmargin = 5mm]
    \item In this work, we propose a novel continual user representation learning method, called \proposed, that 1) retains the learning capability until the end of the training sequence,
    2) captures the relationship between tasks.
    and, 3) prevents catastrophic forgetting through a novel knowledge retention module with pseudo-labeling.
    
    \item For an efficient training of~\proposed, we propose a relation-aware user sampling strategy that samples users from each task considering the relationship between tasks. 
    
    \item Extensive experiments on two public and one proprietary datasets demonstrate the superiority of \proposed~compared to the recent state-of-the-art methods.
    A further appeal of \proposed~is its robustness in real-world scenarios, verifying the possibility of adopting \proposed~on various web platforms.
\end{itemize}

\section{Related Work}
\subsection{Universal User Representation}
User modeling (UM) refers to the process of obtaining the user profile, which is a conceptual understanding of the user for personalized recommender systems.
The key idea of UM is to learn the representation for each user by leveraging the user's interacted items or the features of the items, and the obtained representations are used for a wide range of applications such as response prediction and recommendation \cite{10.5555/3491440.3492135}.
Early methods aim to learn a user representation via matrix factorization, whose matrix consists of user-item interaction history, assuming that a user can be represented based on the interacted items \cite{10.1145/1242572.1242643, 10.5555/2981562.2981720}.
Along with the recent advances of deep neural networks, neural network based user modeling got a surge of interest from researchers, including factorization-based approaches \cite{guo2017deepfm, 10.1145/3077136.3080777},
recurrent neural network based approaches \cite{hidasi2015session},
and graph-based approaches \cite{wang2019neural,wu2019session}.
However, these UM approaches focus on task-specific user representations, which may not represent the generalized interest of users.

Recently, several studies have adopted multi-task learning (MTL) or transfer learning (TL) to move from task-specific user representations to universal user representations.
MTL aims to simultaneously learn multiple tasks with a single shared representation for each user.
For example, DUPN \cite{ni2018perceive} introduces an attention mechanism to integrate all content, behavior, and other relevant information from multiple tasks to generate a user representation.
ESM$^2$ \cite{10.1145/3397271.3401443} utilizes the conditional probability of user behavior graph (e.g., impression $\rightarrow$ click $\rightarrow$ purchase) to explicitly express the task relation for multi-task learning.
On the other hand, TL exploits the knowledge (i.e., user representation) gained in a source task, which is different but positively related to the target task, to improve the performance of target task \cite{pmlr-v97-houlsby19a,10.5555/2900423.2900630}.
For example, DARec \cite{yuan2019darec} models the rating pattern in the source task and transfers the knowledge to the target task for the cross-domain recommendation.
Moreover, PeterRec \cite{yuan2020parameter} boosts the model performance on various target tasks by pre-training the model with sequential user-item interactions in a source task in a self-supervised manner.

Despite the success of MTL and TL in certain applications, they also have limitations when it comes to real-world scenarios.
More precisely, MTL requires all the tasks and their associated data to be available in advance, which is problematic when a new service is to be launched as the model should be retrained with all the data.
Moreover, TL can be done with only two tasks, i.e., the source task and the target task, which is impractical for real-world online platforms that contain multiple target tasks.
Distinguished from MTL and TL, we propose to continually learn universal user representations, which is more practical in reality.

 
\subsection{Continual Learning}
It is widely known that neural networks tend to forget the knowledge of previously learned tasks when they are trained on a new task, and this phenomenon is called catastrophic forgetting \cite{mccloskey1989catastrophic,ratcliff1990connectionist}.
Continual learning aims to prevent such catastrophic forgetting issues during the training of a stream of various tasks.
Recently proposed continual learning approaches can be divided into three categories, i.e., replay-based, architecture-based, and parameter regularization-based approaches.
\textbf{Replay-based approaches} prevent catastrophic forgetting by storing a subset of data of previous tasks into a replay buffer, and use them when training on the next task.
Therefore, selecting the subset of data that best represents the data distribution in the previous tasks is the key to the success of replay-based approaches \cite{yoon2021online,kumariretrospective,shin2017continual}.
\textbf{Architecture-based approaches} prevent catastrophic forgetting by expanding the capacity of the model architecture when it is insufficient to train on new tasks \cite{yoon2017lifelong,hung2019compacting}.
\textbf{Regularization-based approaches} regularize the model parameters to minimize catastrophic forgetting. 
The key idea is to restrict significant changes in the model parameters during training on new tasks by regularizing the model parameters that were important in previous tasks \cite{kirkpatrick2017overcoming,mallya2018piggyback}.
Recently, PackNet \cite{mallya2018packnet} leverages binary masks to restrict the update of the parameters that were shown to be important in previous tasks.

Inspired by the success of PackNet, CONURE \cite{yuan2021one} proposes to learn the universal user representation by keeping the parameters that were crucial in previous tasks.
Specifically, CONURE continuously learns the sequence of tasks by iteratively removing less important parameters and saving the crucial parameters for each task, which is also called network pruning.
By freezing the crucial parameters in previous tasks, CONURE is able to learn new tasks while retaining the knowledge obtained from the previous tasks, and such an approach is called parameter isolation.
However, such parameter isolation-based methods restricts the modification of model parameters that are used in previous tasks, and thus the model's learning capacity is gradually reduced as new task are introduced due to the lack of available learnable parameters.
Moreover, the knowledge obtained from the new tasks cannot be transferred to the previous tasks, i.e., backward transfer, which can be also helpful for previous tasks.


\section{Preliminaries}
In this section, we introduce a formal definition of the problem including the notations and the task description (Sec.~\ref{Sec: Problem Formuation}).
Then, we briefly introduce NextitNet \cite{yuan2019simple}, which is used as the backbone network of our model (Sec.~\ref{Sec: Model Architecture: NextitNet}), and
how the backbone network is trained in the sequence of tasks $\mathcal{T}$ (Sec.~\ref{Sec: Task Incremetal Learning}).

\subsection{Problem Formulation}
\label{Sec: Problem Formuation}
\noindent \textbf{Notations.}
Let $\mathcal{T} = \left\{T_1, T_2, \cdots, T_{M}\right\}$ denote the set of consecutive tasks, which can be also represented as $T_{1:M}$.
Let $\mathcal{U} = \left\{u_1, u_2, \cdots, u_{N}\right\}$ denote the set of users. 
Each user $u_l\in\mathcal{U}$ is represented by his/her behavior sequence $\mathbf{x}^{u_l} = \left\{x_1^{u_l},x_2^{u_l},\cdots,x_{n}^{u_l}\right\}$, where $x_t^{u_l}\in\mathcal{I}$ is the $t$-th interaction of $u_l$ and $\mathcal{I}$ is the set of items.
For each task $T_i$, only a subset of users $\mathcal{U}^{T_i} = \{u_1, u_2, \cdots, u_{|\mathcal{U}^{T_i}|}\}$ exist, i.e., $\mathcal{U}^{T_i} \subset \mathcal{U}$,
and the set of users in task $T_i$ is associated with the set of labels $\mathbf{Y}^{T_i} = \{y^{T_i}_{u_1}, \ldots, y^{T_i}_{u_{|\mathcal{U}^{T_i}|}} \}$, where $y^{T_i}_{u_l}$ denotes the label of user $u_l$ in task $T_i$ (e.g., purchased item, gender, and age) of user $u_l$. We use $\mathcal{Y}^{T_i}$ to denote the set of unique labels in $\mathbf{Y}^{T_i}$, and $\mathbf{y}_{u_l}^{T_i}\in\{0,1\}^{|\mathcal{Y}^{T_i}|}$ denotes the one-hot transformation of $y^{T_i}_{u_l}$. 
Note that the first task $T_1$ contains all the users in the dataset, i.e., $\mathcal{U}^{T_1} = \mathcal{U}$.

\smallskip
\noindent \textbf{Task: Continual User Representation Learning.}
Assume that we are given the set of consecutive tasks $\mathcal{T} = \left\{T_1, T_2, \cdots, T_{M}\right\}$, where each task $T_i$ is associated with the set of users $\mathcal{U}^{T_i}$ and the set of labels $\mathbf{Y}^{T_i}$. 
Our goal is to train a single model $\mathcal{M}$ on each task $T_i\in\mathcal{T}$ one by one in a sequential manner to predict the label of each user $u_l\in\mathcal{U}^{T_i}$, i.e., $\mathbf{y}_{u_l}^{T_i} = G^{T_i}(\mathcal{M}(\mathbf{x}^{u_l}))$, where $\mathcal{M}$ is the backbone encoder network that generates universal user representations, and $G^{T_i}$ is a task-specific classifier for task $T_i$. After training the entire sequence of tasks from $T_1$ to $T_M$, the single model $\mathcal{M}$ is used to serve all tasks in $\mathcal{T}$.

\subsection{Model Backbone: TCN}
\label{Sec: Model Architecture: NextitNet}
\looseness=-1
Following a previous work \cite{yuan2021one}, we adopt temporal convolutional network (TCN) \cite{yuan2019simple} as the backbone network of \proposed, although our framework is network-agnostic.
TCN learns the representation of a user $u_l$ based on the sequence of the user's interacted items, i.e., $\mathbf{x}^{u_l} = \left\{x_1^{u_l},x_2^{u_l},\cdots,x_{n}^{u_l}\right\}$. 
More precisely, given an initial embedding matrix of $\mathbf{x}^{u_l}$, i.e., $\mathbf{E}^{u_l}_0\in\mathbb{R}^{n\times f}$ where $f$ is the emebdding size, we pass it through a TCN, which is composed of a stack of residual blocks, each containing two temporal convolutional layers and normalization layers. The $k$-th residual block is given as follows:
\begin{equation}
\small
    \mathbf{E}^{u_l}_k = {F}_k(\mathbf{E}^{u_l}_{k-1}) + \mathbf{E}^{u_l}_{k-1} = R_k(\mathbf{E}^{u_l}_{k-1})
    \label{Eq 1}
\end{equation}
where $k=1,\dots,K$, $\mathbf{E}^{u_l}_{k-1}$ and $\mathbf{E}^{u_l}_k$ are the input and output of the residual block in layer $k$, respectively, ${F}_k$ is a residual mapping in layer $k$, which is composed of two layers of convolution, layer normalization and ReLU activation, and ${R}_k$ is the residual block in layer $k$.

\subsection{Task Incremental Learning}
\label{Sec: Task Incremetal Learning}
\looseness=-1
During training, our backbone network sequentially learns the tasks in $\mathcal{T}$ one by one, i.e., $T_1 \rightarrow T_2 \rightarrow \cdots \rightarrow T_M$.
Following \cite{yuan2021one}, 
the same user behavior sequence $\mathbf{x}^{u_l}$ is used as the model input for all tasks, 
whereas the target domain $y_{u_l}^{T_i}$ varies as the task varies, e.g., item purchase prediction for $T_1$, user age prediction for $T_2$, and user gender prediction for $T_3$, etc.

\smallskip
\noindent \textbf{Training the First Task ($T_1$)}. 
In the first task $T_1$, we generate base user representations, which will be used in the following tasks, based on the behavior sequence of user $u_l\in\mathcal{U}^{T_1}$, i.e., $\mathbf{x}^{u_l}$.
To do so, we train our backbone network in a self-supervised manner to autoregressively predict the next item in the user behavior sequence $\mathbf{x}^u$ as follows:
\begin{equation}
\small
    p(\mathbf{x}^{u_l};\Theta)=\prod_{j=1}^{n-1} p(x_{j+1}^{u_l}|x_1^{u_l},\cdots,x_{j}^{u_l}; \Theta),
    \label{Eq 3}
\end{equation}
where $p(x_{j+1}^{u_l}|x_1^{u_l},\cdots,x_{j}^{u_l}; \Theta)$ indicates probability of the $(j+1)$-th interaction with user $u_l$ conditioned on the user's past interaction history $\{x_1^{u_l},\cdots,x_{j}^{u_l}\}$, and $\Theta$ is the set of parameters of TCN.
By training TCN to maximize the above joint probability distribution of the user behavior sequence $\mathbf{x}^{u_l}$, we obtain the base user representation for user $u_l$, which can be transferred to various tasks $T_{2:i}$.

\smallskip
\noindent \textbf{Training of $T_{i}$ $(i>1)$.} 
After training the first task $T_1$, we continually learn the subsequent tasks $T_{2:i}$ with our backbone network whose parameters are pre-trained based on $T_1$. 
Given the behavior sequence of user $u_l$, i.e., $\mathbf{x}^{u_l} = \{x_1^{u_l},x_2^{u_l},\cdots,x_{n}^{u_l}\}$, our backbone network autoregressively outputs the embeddings, i.e., $\mathbf{E}_K^{u_l}\in\mathbb{R}^{n\times f}$, for predicting the next item. More precisely, the $j$-th row of $\mathbf{E}_K^{u_l}$ is the output of the network when the behavior sequence $\{x_1^{u_l}, x_2^{u_l}, \cdots, x_{j-1}^{u_l}\}$ is given. Hence, we use the last row of $\mathbf{E}_K^{u_l}$ as the input to the task-specific classifier of $T_i$ to obtain the label (e.g., predicting the next purchased item, gender, age, etc) for user $u_l$ as follows:
\begin{equation}
\small
    \mathbf{\hat{y}}^{T_i}_{u_l} = \mathbf{E}_K^{u_l}[-1,:]\mathbf{W}^{T_i} + \mathbf{b}^{T_i} = G^{T_i}(\mathbf{E}_K^{u_l})
\end{equation}
where $\mathbf{\hat{y}}^{T_i}_{u_l}\in\mathbb{R}^{|\mathcal{Y}^{T_i}|}$ is the prediction of labels of $u_l$ in task $T_i$, $\mathbf{E}_K^{u_l}[-1,:]\in\mathbb{R}^{f}$ is the last row of $\mathbf{E}^{u_l}_K$, $\mathbf{W}^{T_i} \in \mathbb{R}^{f\times |\mathcal{Y}^{T_i}|}$ and $\mathbf{b}^{T_i} \in \mathbb{R}^{|\mathcal{Y}^{T_i}|}$ denote the task-specific projection matrix and bias term of fully-connected-layer for $T_i$, respectively, and $G^{T_i}$ is a simplified notation of the task-specific classifier of $T_i$.


\subsection{Parameter Isolation}
\label{Sec: Parameter Isolation}
However, as shown in \cite{yuan2021one} the model that is naively trained with various tasks in a sequential manner suffers from catastrophic forgetting \cite{mccloskey1989catastrophic,ratcliff1990connectionist}.
That is, the model performance on the previous tasks $T_{1:(i-1)}$ drastically deteriorates when learning on a new task $T_i$. 
To this end, CONURE \cite{yuan2021one} proposes to alleviate catastrophic forgetting by adopting the parameter isolation approach, which freezes a portion of model parameters that were crucial in learning previous tasks. 
Specifically, CONURE learns the new task $T_i$ with the following model parameters:
\begin{equation}
\small
    \mathbf{Z}_k^{T_i} = \textsf{stop\_grad}\left[\mathbf{\hat{Z}}_{k}^{T_{i-1}}\odot \sum_{t=1}^{i-1}\mathbf{M}_k^{T_t}\right]+\mathbf{\hat{Z}}_{k}^{T_{i-1}}\odot (\mathbf{1}_{k} - \sum_{t=1}^{i-1}\mathbf{M}_k^{T_t})
    \label{Eq 4}
\end{equation}
where $\odot$ is element-wise product operator, $\textsf{stop\_grad}$ is an operator that prevents back propagation of gradients, $\mathbf{\hat{Z}}_k^{T_{i-1}}\in\mathbb{R}^{a\times b}$ is the parameter of the $k$-th layer of TCN that is used for learning $T_{i-1}$, where $a\times b$ is the shape of the parameter\footnote{Although the convolution layer in TCN has 3D filters, we reshape it to $a\times b$ for simplicity of explanation.}, and $\mathbf{1}_k\in\mathbb{R}^{a\times b}$ is a matrix whose elements are all ones. Moreover, $\mathbf{M}_k^{T_t}\in\mathbb{R}^{a\times b}$ indicates a task-specific binary mask in layer $k$ for protecting model parameters that were considered to be significant in previous tasks $T_{1:(i-1)}$,
which is obtained as follows:
\begin{equation}
\small
    \centering
    \mathbf{M}_k^{T_i} = \begin{cases}
    1 &\mbox{if } |\mathbf{Z}_k^{T_i}|\odot (\mathbf{1}_{k} - \sum_{t=1}^{i-1}\mathbf{M}_k^{T_t})> \delta,\\
    0 &\mbox{otherwise }
    \end{cases}
    \label{eq:mask_conure}
\end{equation}
where $|\cdot |$ denotes the symbol of element-wise absolute value, and $\delta$ is a threshold hyper-parameter for selecting the significant parameters, which is determined in advance according to the amount of parameters to be frozen in each task. 
In other words, Equation~\ref{eq:mask_conure} aims to keep an element of the parameter $\mathbf{Z}_k^{T_i}$ frozen, if its absolute value is greater than a threshold $\delta$.
In the inference phase, given a $T_t$, the model loads the masks that are stored to generate the representation, i.e., $\mathbf{\tilde{Z}}_k^{T_t} = \mathbf{\hat{Z}}_k^{T_t}\odot\sum_{j=1}^t \mathbf{M}_k^{T_j}$.
In the inference phase, since CONURE utilizes the parameters that were kept frozen after learning each task, it prevents catastrophic forgetting.

\smallskip
\noindent\textbf{Limitation of CONURE. }
However, we argue that CONURE has the following drawbacks:
\textbf{1)} 
By adopting the parameter isolation approach, the number of available learnable parameters is gradually reduced as more tasks are continually added, which limits the model’s capability in learning subsequent tasks even making it impossible to learn new tasks after using up all the remaining parameters. 
Moreover, once the parameters are frozen, the previous tasks cannot benefit from the subsequent tasks whose knowledge can also be beneficial to previous tasks, i.e., positive backward transfer.
\textbf{2)} Besides the limitations incurred by the parameter isolation, CONURE does not consider the relationship between  tasks, which however is beneficial in two perspectives.
That is, it encourages the knowledge learned from positively related tasks to be better transferred between the one another, and at the same time, it prevents the knowledge learned from the negatively related task from disturbing one another.
Therefore, we need a model that learns universal user representations that a) retains the learning capability until the end of the training sequence, while b) considering the relationship between tasks.


\section{Proposed Method:~\proposed}
In this section, we introduce a novel universal user representation learning model, called \proposed, whose key component is the \textit{task embedding}, which not only generates a task-specific soft mask, but also facilitates the relationship between the tasks to be captured.
\begin{figure}[t]
    \centering
    \includegraphics[width=1.0\linewidth]{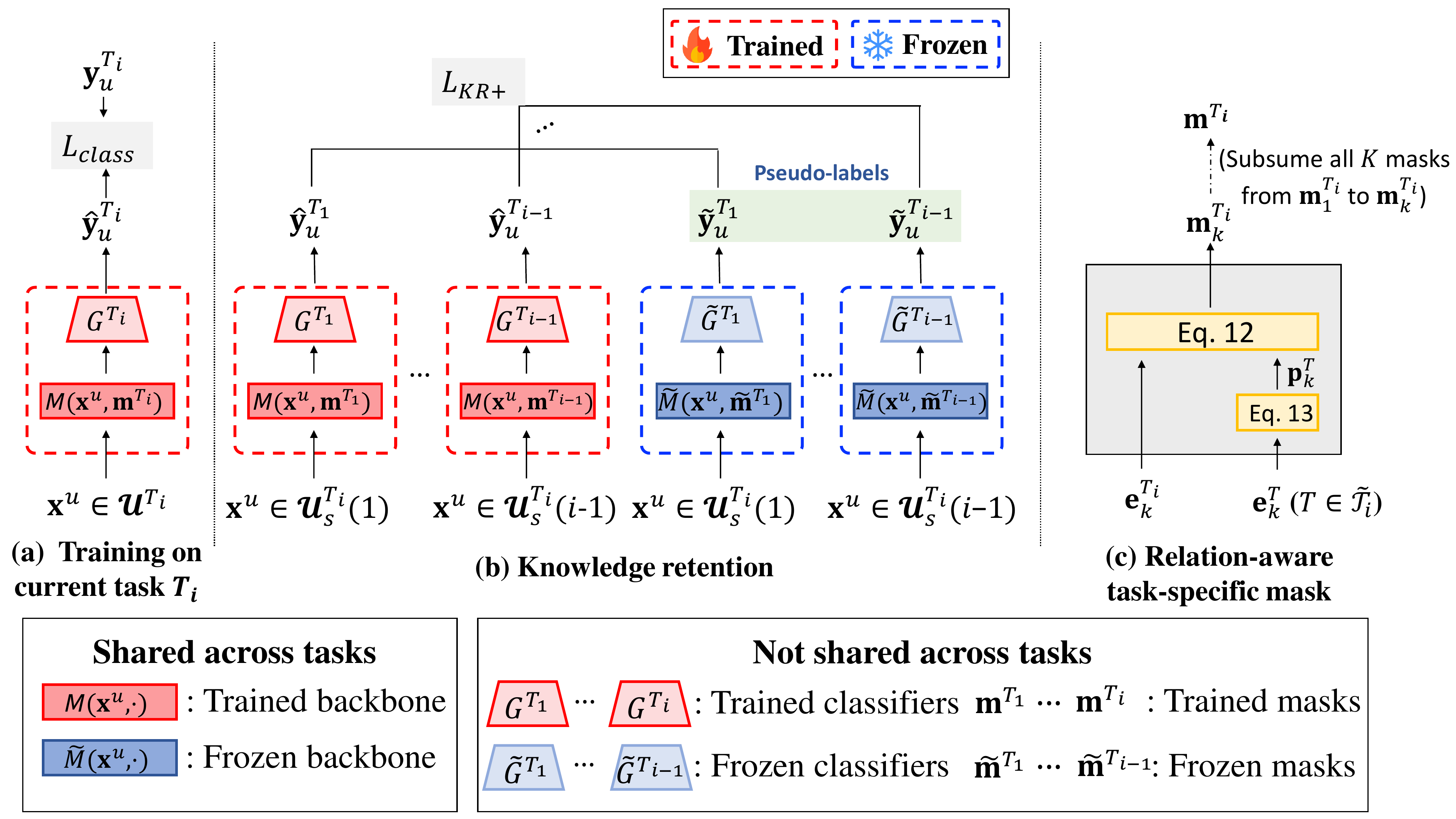}
    \vspace{-1ex}
    \caption{Overall model framework. (a) Training the current task $T_i$. (b) Knowledge retention from previous tasks $T_{1:(i-1)}$. (c) Generating relation-aware task-specific mask $\boldsymbol{m}^{T_i}$.}
    \label{fig: framework}
\end{figure}
\subsection{Learning Task-specific Mask via Task Embedding}
\label{Sec: Learning Relation-aware Mask}
We begin by describing how the task embedding and the task-specific mask are defined.
We use $\mathbf{e}_{k}^{T_i}\in\mathbb{R}^f$ and $\mathbf{m}_{k}^{T_i}\in\mathbb{R}^f$ to denote the task embedding and the task-specific mask of task $T_i$ in layer $k$, respectively, which is defined as follows:
\begin{equation}
\small
    \mathbf{m}_{k}^{T_i} = \sigma (s \cdot \mathbf{e}_k^{T_i})
\label{eq:mask_ours}
\end{equation}
where $\sigma$ is the sigmoid function, and $s$ is a positive scaling hyper-parameter, which determines how much to amplify (or suppress) the layer output of significant (or insignificant) during training.
Then, given the behavior sequence of user $u_l$, i.e., $\mathbf{x}^{u_l}$, we use the task-specific mask $\mathbf{m}_{k}^{T_i}$ to obtain the task-specific output of $u_l$ in task $T_i$ as follows:
\begin{equation}
\begin{split}    
\mathbf{E}^{u_l}_k={F}_k(\mathbf{E}^{u_l}_{k-1};\mathbf{m}_k^{T_i})+ \mathbf{E}^{u_l}_{k-1} ={R}_k(\mathbf{E}^{u_l}_{k-1};\mathbf{m}_k^{T_i})
\end{split}
\end{equation}
where ${F}_k(\mathbf{E}^{u_l}_{k-1};\mathbf{m}_k^{T_i})$ is the masked version of the residual mapping in layer $k$, obtained by element-wise multiplying the mask $\mathbf{m}_k^{T_i}$ (for convenience, we omit the notation of the layer indices in the residual block of the mask.) with each output of the ReLU activation in $F_k(\cdot)$, while $R_k$ represents the residual block in layer $k$.
We use $\mathcal{M}$ to denote the complete backbone network as well as the task embeddings from layer 1 to $K$, which is the single model we aim to learn to serve all tasks. The final output embedding of user $u_l$ in task $T_i$, i.e., $\mathbf{E}^{u_l}_K$, is formulated as follows:
\begin{equation}
\small
    \mathbf{E}^{u_l}_K=\mathcal{M}(\mathbf{x}^{u_l};\mathbf{m}^{T_i})
\end{equation}
where $\mathbf{m}^{T_i}$ is the task-specific mask of task $T_i$ with a slight abuse of notation\footnote{$\mathbf{m}^{T_i}$ subsumes all $K$ masks from $\mathbf{m}_1^{T_i}$ to $\mathbf{m}^{T_i}_K$}.
Note that although the user ID is given for each user $u_l\in\mathcal{U}$, it is not used to obtain $\mathbf{E}^{u_l}_K$.

\looseness=-1
Our proposed masking strategy defined in Equation~\ref{eq:mask_ours} is different from that of CONURE defined in Equation~\ref{eq:mask_conure} in two ways:
(1) The mask of CONURE aims to freeze a small portion of parameters after having trained on a task, which gradually reduces the number of remaining available learnable parameters as more tasks are continually added. On the other hand, the mask of our proposed method does not freeze any of the parameters, thereby retaining the learning capability until the end of the training sequence. Moreover, we define our mask to be a \textit{soft} continuous mask to allow all the parameters to be used for learning every task, whereas the mask of CONURE is a \textit{hard} binary mask.
(2) The mask of CONURE is applied on the actual model parameters in each layer (e.g., weights of convolutional filters in a CNN layer), whereas our mask is applied on the output of each layer. The major difference is in the size of the masks, i.e., the mask of CONURE is greatly larger than our mask. The reduced size of the mask not only enhances the model efficiency, but also enables our proposed method to better capture the relationship between tasks by obtaining a more compact representation of each task, i.e., $\mathbf{m}_{k}^{T_i}$.

\subsection{Overcoming Catastrophic Forgetting via Knowledge Retention
}
\label{Sec: Knowledge Retention via Pseudo Labeling}
However, as \proposed~is not based on the parameter isolation approach, simply applying the masking strategy as described in Section~\ref{Sec: Learning Relation-aware Mask} is prone to catastrophic forgetting, because we allow the parameters to be shared across tasks instead of freezing parameters used in previous tasks.
Hence, to prevent catastrophic forgetting, we propose to transfer the knowledge of the current model regarding previous tasks to help train the current model itself. 
The key idea is, while training task $T_i$, to utilize the current backbone network $\mathcal{M}$, and $i-1$ task-specific classifiers of previous tasks, i.e., $G^{T_{1:(i-1)}}$, to generate the pseudo-labels for user $u_l\in\mathcal{U}^{T_i} \subset \mathcal{U}$ as follows:
\begin{equation}
\small
\mathbf{\tilde{y}}^{T_j}_{u_l} = \tilde{G}^{T_j}(\mathcal{\tilde{M}}(\mathbf{x}^{u_l};\mathbf{\tilde{m}}^{T_j})) \quad\text{for} \quad j=1,\dots,i-1
\end{equation}
where $\mathbf{\tilde{y}}^{T_j}_{u_l}\in\mathbb{R}^{|\mathcal{Y}^{T_j}|}$ is the pseudo-label of user $u_l$ in task $T_j$, and $\tilde{G}$, $\mathcal{\tilde{M}}$, and $\mathbf{\tilde{m}}$ indicate frozen task-specific classifier, backbone network, and task-specific mask, respectively, which are only used to generate the pseudo-labels.
Note that since each user $u_l\in\mathcal{U}^{T_i}\subset \mathcal{U}$ is represented by his/her behavior sequence in every task as described in Sec.~\ref{Sec: Problem Formuation}, we can obtain the pseudo-label for each user in different tasks, even if user $u_l$ is not observed in previous tasks $T_{1:(i-1)}$.
Given the pseudo-labels, we minimize the following loss to retain the knowledge of previous tasks by training the current backbone model $\mathcal{M}$ to predict the pseudo-labels of $u_l$ obtained from previous tasks $T_{1:(i-1)}$:

\begin{equation}
\small
    \mathcal{L}_\text{KR} = \underset{1\leq j < i}{\mathbb{E}}\left[\underset{u_l\in\mathcal{U}^{T_i}}{\mathbb{E}}\left[{L_{\text{MSE}}}({G}^{T_j}(\mathcal{M}(\mathbf{x}^{u_l};\mathbf{m}^{T_j})),\mathbf{\tilde{y}}^{T_j}_{u_l})\right]\right]
\label{Eq 11}
\end{equation}
where ${L}_\text{MSE}$ is the mean squared error loss. That is, we continually update the backbone (i.e., $\mathcal{M}$) as well as the previous task-specific classifiers (i.e., $G^{T_{1:(i-1)}}$) based on the pseudo-labels generated from the backbone and the task-specific classifiers, which are obtained after the training of task $T_{i-1}$ is finished, i.e.,
before we start training on the current task $T_i$.
By doing so,~\proposed~retains the knowledge of the current model regarding the previous tasks, which helps to alleviate catastrophic forgetting.

\subsection{Relation-aware Task-specific Mask}
\looseness=-1
Recall that the second limitation of CONURE is that it fails to capture the relationship between tasks, since parameters used to learn previous tasks are kept frozen while learning new task.
In this section, we describe how the relationship between tasks is captured by using the task-specific mask $\mathbf{m}_{k}^{T_i}$ computed in Equation~\ref{eq:mask_ours}.
Specifically, when training on a new task, we aggregate the information from the previous tasks along with the current new task. 
Let $T_i$ denote the current task, and assume that previous tasks, i.e., $T_{1:{i-1}}$, have been trained by the model. We define an aggregate set of task $T_j$ as follows:
\begin{equation}
\small
\tilde{\mathcal{T}}_j=\left\{T_r|T_r \in T_{1:i}, \text{where} \,\,(T_i=\,\textsf{current task})\,\text{and}\,(j\neq r)\right\}    
\end{equation}
where $\tilde{\mathcal{T}}_j$ is an aggregate set of task $T_j$.
For example, given that we have already trained on $T_1$ and $T_2$, and the current task to be trained is the new task $T_3$, the aggregate sets for each task is given as: $\tilde{\mathcal{T}}_1: \{T_2, T_3\}$, $\tilde{\mathcal{T}}_2: \{T_1, T_3\}$, $\tilde{\mathcal{T}}_3: \{T_1, T_2\}$.
Given the aggregate set of each task $T_i$, we update the task-specific mask $\mathbf{m}_k^{T_i}$ computed in Equation~\ref{eq:mask_ours} to obtain the \textit{relation-aware task-specific mask} as follows:
\begin{equation}
\small
    \mathbf{m}_{k}^{T_{i}} = \sigma\left(s\cdot f_{k}^{T_i}\left[\text{tanh}(s\cdot \mathbf{e}_{k}^{T_i})\parallel(\parallel_{T \in \tilde{\mathcal{T}}_i} \mathbf{p}_{k}^{T})\right]\right)\in\mathbb{R}^f
    \label{eq:relation_aware}
\end{equation}
where $\parallel$ is the vertical concatenation operation, $f_{k}^{T_i}$ is a 1-layer MLP, i.e., $\mathbb{R}^{(2|\tilde{\mathcal{T}}|+1)\times f} \rightarrow \mathbb{R}^{f}$, which maps the concatenated task embeddings to a new task embedding of $T_i$, and $\mathbf{p}_{k}^{T}$ is defined as follows:
\begin{equation}
\small
    \mathbf{p}_{k}^{T} = \left[\text{tanh}(s \cdot\mathbf{e}_{k}^{T})\parallel \text{tanh}(-s\cdot \mathbf{e}_{k}^{T})\right]\in\mathbb{R}^{2\times f}
    \label{eq:p_kT}
\end{equation}
\looseness=-1
Recall that the task embedding $\mathbf{e}^{T}$ contains information about how much to amplify (or suppress) the layer output in task $T$. To provide a more nuanced understanding of task $T$, instead of directly using $\text{tanh}(s\cdot\mathbf{e}_{k}^{T})$, we introduce additional $\text{tanh}(-s\cdot\mathbf{e}_{k}^{T})$, which provides information about the opposite relatedness with $\text{tanh}(s\cdot\mathbf{e}_{k}^{T})$. Hence, $f_k^{T_i}$ learns to distinguish between the positive and negative information of tasks in $\mathcal{\tilde{T}}_i$. Moreover, it learns to determine which dimension of $\left[\text{tanh}(s\cdot \mathbf{e}_{k}^{T_i})\parallel(\parallel_{T \in \tilde{\mathcal{T}}_i} \mathbf{p}_{k}^{T})\right]$ should send amplification or suppression signals to $\mathbf{m}_k^{T_i}$.
In this situation, $\text{tanh}(-s\cdot\mathbf{e}_{k}^{T})$ serves as a counterbalance to the $\text{tanh}(s\cdot\mathbf{e}_{k}^T)$, allowing  $f_k^{T_i}$ to amplify the information of specific tasks (i.e., $T$), but also suppressing the same task if necessary. Therefore, this methods serves to provide a more complete and subtle fine-tune of representation for the new task embeddings.
In summary, we obtain a relation-aware task-specific mask for task $T_i$, i.e., $\mathbf{m}_k^{T_i},$ by encoding the information regarding $\mathcal{\tilde{T}}_i$.

\subsection{Relation-aware User Sampling Strategy}
\label{Sec: Relation-aware User Sampling}

However, generating pseudo-labels and optimizing the loss $\mathcal{L}_\text{KR}$ defined in Equation~\ref{Eq 11} for each previous task using all the training data given in the current task is memory-inefficient and time-consuming. 
To alleviate the complexity issues, instead of using all the users in the current task, we propose to sample users from the current task that will be involved in the knowledge retention process.
The key assumption is that retaining knowledge of a previous task would be easier if the current task is more similar to the previous task.
Hence, we propose a sampling strategy that samples less users from the current task when addressing a previous task that is more similar to the current task. 
More precisely, given the current task $T_i$ and a previous task $T_j\in T_{1:(i-1)}$, we sample a subset of users $\mathcal{U}_{s}^{T_i}$ uniformly at random from $\mathcal{U}^{T_i}$ as follows:
\begin{equation}
\small
    \mathcal{U}_{s}^{T_i}(j) \leftarrow \textsf{sample}(\mathcal{U}^{T_i},\rho_{i,j})
    \label{eq: rho}
\end{equation}
where $\rho_{i,j}$ is the sampling rate that determines the amount of samples. Specifically, $(\rho_{i,j}\times 100)$\% of users are sampled from $\mathcal{U}^{T_i}$, and $\rho_{i,j}$ is defined as follows:
\begin{equation}
\small
    \rho_{i,j} = 1-\frac{1}{K}\sum^K_{k=1}\sigma(c\times\text{cos}(\mathbf{m}_{k}^{T_i},{\mathbf{\tilde{m}}}_{k}^{T_j}))
    \label{Eq:sampling}
\end{equation}
\looseness=-1
where $\text{cos}(\cdot,\cdot)$ is cosine similarity, $c$ is a scaling hyper-paramter, and ${\mathbf{\tilde{m}}}_{k}^{T_j}$ is the frozen mask of task $T_j$ that is generated from the model before we start training the current task $T_i$. 
The main idea is to assign a small $\rho_{i,j}$ to a previous task $T_j$, if the similarity between $T_j$ and the current task $T_i$ is high, which encourages the model to sample a small number of users from similar tasks, while a large number of users are sampled from dissimilar tasks. Given sampled subsets of users for the current task $T_i$, we optimize the following loss:
\begin{equation}
\small
    \mathcal{L}_\text{KR+} = \underset{1\leq j < i}{\mathbb{E}}    
    \left[  \frac{\rho_{i,j}}{\sum_{k=1}^{i-1}\rho_{i,k}}\sum_{u_l\in\mathcal{U}_{s}^{T_i}(j)}\mathcal{L_{\text{MSE}}}({G}^{T_j}(\mathcal{M}(\mathbf{x}^{u_l};\mathbf{m}^{T_j})),\mathbf{\tilde{y}}^{T_j}_{u_l})\right]
\label{Eq 16}
\end{equation}
By doing so, our model retains the knowledge from similar tasks efficiently and dissimilar tasks with a small amount of samples.
We empirically observe in our experiments that the sampling ratio $\rho_{i,j}$ is within the range between 0.007 and 0.08, which implies that~\proposed~is efficiently trained in terms of both time and memory.

\subsection{Training and Inference}
\subsubsection{Training}
For a given task $T_i$, \proposed~is trained by optimizing the following objective function:
\begin{equation}
\small
    \mathcal{L} = \mathcal{L}_\text{class} + \alpha \mathcal{L}_\text{KR+}
    \label{Eq 17}
\end{equation}
where $\alpha$ is a coefficient that controls the contribution of the knowledge retention module. The classification loss, i.e., $\mathcal{L}_\text{class}$, is given as follows:
\begin{equation}
\small
    \mathcal{L}_\text{class} = \underset{u_l\in\mathcal{U}^{T_i}}{\mathbb{E}}\left[{L_\text{CE}}(G^{T_i}(\mathcal{M}(\mathbf{x}^{u_l};\mathbf{m}^{T_i})),\textbf{{y}}_{u_l}^{T_i})\right],
\label{Eq 18}
\end{equation}
where $L_\text{CE}$ is the cross-entropy loss, and $\mathbf{{y}}^{T_i}_{u_l}\in\{0,1\}^{|\mathcal{Y}^{T_i}|}$ is the ground truth one-hot label vector of user $u_l$ in task $T_i$. Note that the above optimization is done for all $M$ tasks in $\mathcal{T}$ in a sequential manner, i.e., $T_1 \rightarrow T_2 \rightarrow \cdots \rightarrow T_M$.

\smallskip
\noindent\textbf{A Training Trick: Annealing $s$. }
In Eq.~\ref{eq:mask_ours}, when the positive scaling hyper-parameter $s$ goes to infinity, i.e., $s \rightarrow \infty$, the task-specific mask operates like a step function i.e., $\mathbf{m}^{T_i}\rightarrow \left\{0,1\right\}^f$. 
This makes the gradient to flow into only a small portion of the task embedding $\mathbf{e}^{T_i}$ during training, which prevents an efficient training.
To ensure a proper gradient flow on all task embeddings in the course of training, we introduce an annealing strategy \cite{serra2018overcoming} on $s$:
\begin{equation}
\small
    s =\frac{1}{s_\text{max}} + (s_\text{max} - \frac{1}{s_\text{max}})\frac{b-1}{B-1}
    \label{Eq 19}
\end{equation}
where $b$ is the batch index, $B$ is the total number of batches in an epoch, and $s_\text{max}$ is a positive scaling hyper-parameter. 
We generate masks with $s$ during training and $s_\text{max}$ during inference.
\subsubsection{Inference}
Having trained all the tasks in $\mathcal{T}$, we perform inference on all the tasks in $\mathcal{T}$. Specifically, given a task $T_i\in\mathcal{T}$, we obtain the task-specific mask $\mathbf{m}^{T_i}$ based on the task embeddings as described in Eq.~\ref{eq:relation_aware} and a task-specific classifier $G^{T_i}$, while fixing $s = s_\text{max}$ in Eq.~\ref{Eq 19}.
Then, 
for each task $T_i\in\mathcal{T}$ and user $u_l\in\mathcal{U}^{T_i}$ whose behavior sequence is given by $\mathbf{x}^{u_l}$, we make the model predictions based on the task-specific user representation generated by utilizing the backbone network $\mathcal{M}$ as follows: 
\begin{equation}
\small
 \mathbf{\hat{y}}_{u_l}^{T_i}=G^{T_i}(\mathcal{M}(\mathbf{x}^{u_l};\mathbf{m}^{T_i})).
\end{equation}
where $\mathbf{\hat{y}}^{T_i}_{u_l}\in\mathbb{R}^{|\mathcal{Y}^{T_i}|}$ is the prediction of labels of $u_l$ in task $T_i$.

\begin{table}[t]
\centering
\caption{Data Statistics ($|\mathcal{U}^{T_i}|$: num. users in $T_i$, $|\mathcal{Y}^{T_i}|$: num. unique labels in $T_i$).}
\vspace{-2ex}
\renewcommand{\arraystretch}{1.1}
    \resizebox{0.99\linewidth}{!}{
    \begin{tabular}{wc{1cm}|wc{0.8cm}wc{0.5cm}|wc{0.8cm}wc{0.5cm}|wc{0.8cm}wc{0.5cm}|wc{0.8cm}wc{0.5cm}|wc{0.8cm}wc{0.5cm}|wc{0.8cm}wc{0.5cm}}
    \toprule
     \multirow{2}{*}{Dataset} & \multicolumn{2}{c|}{Task 1 ($T_1$)} & \multicolumn{2}{c|}{Task 2 ($T_2$)} & \multicolumn{2}{c|}{Task 3 ($T_3$)} & \multicolumn{2}{c|}{Task 4 ($T_4$)} & \multicolumn{2}{c|}{Task 5 ($T_5$)} & \multicolumn{2}{c}{Task 6 ($T_6$)}   \\ \cline{2-13}
    & \multicolumn{1}{c}{$|\mathcal{U}^{T_1}|$} & \multicolumn{1}{wc{0.5cm}|}{$|\mathcal{Y}^{T_1}|$} & \multicolumn{1}{c}{$|\mathcal{U}^{T_2}|$} & \multicolumn{1}{wc{0.5cm}|}{$|\mathcal{Y}^{T_2}|$} & \multicolumn{1}{c}{$|\mathcal{U}^{T_3}|$} & \multicolumn{1}{wc{0.5cm}|}{$|\mathcal{Y}^{T_3}|$} & \multicolumn{1}{c}{$|\mathcal{U}^{T_4}|$} & \multicolumn{1}{wc{0.5cm}|}{$|\mathcal{Y}^{T_4}|$} & \multicolumn{1}{c}{$|\mathcal{U}^{T_5}|$} & \multicolumn{1}{wc{0.5cm}|}{$|\mathcal{Y}^{T_5}|$} & \multicolumn{1}{c}{$|\mathcal{U}^{T_6}|$} & \multicolumn{1}{wc{0.5cm}}{$|\mathcal{Y}^{T_6}|$} \\ \midrule \midrule
    \multirow{2}{*}{TTL} & \multicolumn{2}{c|}{Watching} & \multicolumn{2}{c|}{Clicking} & \multicolumn{2}{c|}{Thumb-up} & \multicolumn{2}{c|}{Age} & \multicolumn{2}{c|}{Gender} & \multicolumn{2}{c}{Life status} \\ 
     & \multicolumn{1}{wc{1cm}}{1.47M} & \multicolumn{1}{c|}{0.64M} & 1.39M   & \multicolumn{1}{c|}{17K} & 0.25M & \multicolumn{1}{wc{1cm}|}{7K} & 1.47M & \multicolumn{1}{c|}{8} & 1.46M & \multicolumn{1}{c|}{2} & 1M &  \multicolumn{1}{c}{6}\\ \midrule
    \multirow{2}{*}{ML} & \multicolumn{2}{c|}{Clicking} & \multicolumn{2}{c|}{4-star} & \multicolumn{2}{c|}{5-star} & \multicolumn{2}{c|}{\multirow{2}{*}{-}} & \multicolumn{2}{c|}{\multirow{2}{*}{-}} & \multicolumn{2}{c}{\multirow{2}{*}{-}} \\ 
     & 0.74M & \multicolumn{1}{c|}{54K} & 0.67M & \multicolumn{1}{c|}{26K} & 0.35M &  \multicolumn{1}{c|}{16K} & \multicolumn{2}{c|}{} & \multicolumn{2}{c|}{}   & \multicolumn{2}{c}{} \\ \midrule
    \multirow{2}{*}{\begin{tabular}{@{}c@{}}NAVER \\ Shopping\end{tabular}} & \multicolumn{2}{c|}{Search Query} & \multicolumn{2}{c|}{Search Query} & \multicolumn{2}{c|}{Item Category} & \multicolumn{2}{c|}{Item Category} & \multicolumn{2}{c|}{Gender} & \multicolumn{2}{c}{Age} \\ 
     & {0.9M}  & \multicolumn{1}{c|}{0.58M} & {0.59M} & \multicolumn{1}{c|}{0.51M} & {0.15M} & \multicolumn{1}{c|}{4K} & {0.15M}& \multicolumn{1}{c|}{10} & {0.82M} & \multicolumn{1}{c|}{2} & 0.82M & \multicolumn{1}{c}{9} \\ \bottomrule
    \end{tabular}}
    \vspace{-2ex}
\label{tab: data statistics}
\end{table}

\section{Experiments}

\noindent \textbf{Datasets and Tasks.}
For comprehensive evaluations, we use two public datasets, i.e., Tencent TL (TTL) dataset\footnote{\label{url: TTL}\url{https://drive.google.com/file/d/1imhHUsivh6oMEtEW-RwVc4OsDqn-xOaP/view?usp=sharing}} \cite{yuan2021one,yuan2020parameter} and Movielens (ML) dataset\footnote{\label{url: ml}\url{https://grouplens.org/datasets/movielens/25m/}}, and a proprietary NAVER Shopping dataset.
Since there exists no dataset for continual user representation learning over tasks, we create various tasks for each public dataset following a previous work \cite{yuan2021one}.
The detailed statistics for each dataset are described in Table \ref{tab: data statistics}, and task descriptions are provided as follows:

\begin{itemize}[leftmargin=2mm]
    \item \textbf{TTL dataset}\textsuperscript{\ref{url: TTL}}\cite{yuan2021one,yuan2020parameter} consists of three item recommendation tasks and three user profiling tasks. 
    Specifically, $T_1$ contains userID and the users' recent 100 news \& video watching interactions on the QQ Browser platform. 
    Based on the users' interaction history in QQ Browser platform, $T_2$ and $T_3$ aim to predict clicking interactions and thumb-up interactions on the Kandian platform, respectively.
    Moreover, in $T_4$, $T_5$, and $T_6$, models are trained to predict user's age, gender, and life status, respectively.
    \item \textbf{ML dataset}\textsuperscript{\ref{url: ml}} consists of three tasks, all of which are related to item recommendation.
    Specifically, $T_1$ contains user IDs and their recent 30 clicking interactions, excluding the items that are 4-starred and 5-starred by the user.
    Given the recent 30 clicking interactions, models are trained to predict 4-starred and 5-starred items by the user in $T_2$ and $T_3$.
    \item \textbf{NAVER Shopping dataset} consists of two search query prediction tasks in the portal, two purchased item category prediction tasks in the shopping platform, and two user profiling tasks, which are elaborately designed considering the interests of the real-world industry.
    Specifically, $T_1$ includes userID and the user's recent 60 search queries in the online portal platform. 
    Given the recent 60 search query histories, models are trained to predict next five search queries in $T_2$.
    Moreover, in $T_3$ and $T_4$, models are learn to predict the minor and major categories of user-purchased items based on the search query histories, which can also be recognized as cross-domain recommendations.
    Finally, the models are trained to predict users' gender in $T_5$ and age in $T_6$. We have detailed statistics and descriptions in Table \ref{tab: Ecom Data Statistics} in Appendix \ref{app: A Proprietary Dataset}.
    We argue that obtaining the universal user representation from search queries, which is highly general in nature, is crucial in the online portal platforms since the various services can benefit from the highly transferrable representation.
    To the best of our knowledge, this is the first time the search queries are used to learn universal user representations in continual learning.
\end{itemize}

\begin{table*}[t]
\caption{Overall model performance over various tasks on TTL, ML, and NAVER Shopping datasets.}
\vspace{-2ex}
\resizebox{0.85\linewidth}{!}{
\begin{tabular}{c||cccccc||ccc||cccccc}
\toprule
             &\multicolumn{6}{c||}{TTL} & \multicolumn{3}{c||}{ML} &\multicolumn{6}{c}{NAVER Shopping}\\ \cline{2-16}
             & $T_1$ & $T_2$ & $T_3$ & $T_4$ & $T_5$ & $T_6$ & $T_1$ & $T_2$ & $T_3$ & $T_1$ & $T_2$ & $T_3$ & $T_4$ & $T_5$ & $T_6$\\ \midrule\midrule
SinMo & 0.0446 & 0.0104 & 0.0168 & 0.4475 & 0.8901 & 0.4376 & 0.0566 & 0.0186 & 0.0314 & 0.0349 & 0.0265 & 0.0292 & 0.1984 & 0.5742 & 0.2985 \\ \midrule
FineAll    & 0.0446 & 0.0144 & 0.0218 & 0.5232 & 0.8851 & 0.4596 & 0.0566 & 0.0224 &  0.0328 & 0.0349 & 0.0318 & 0.0332 & 0.2367 & 0.6204 & 0.3247 \\
PeterRec     & 0.0446 & 0.0147 & 0.0224 & 0.5469 & 0.8841 & 0.4749 & 0.0566 & 0.0224 &  0.0308 & 0.0349 & 0.0317 & 0.0322 & 0.2370 & 0.6257 & 0.3258 \\ \midrule
MTL          & -      & 0.0102 & 0.0142 & 0.4672 & 0.8012 & 0.3993 & -      & 0.0144 &  0.0267 & - & 0.0143 & 0.0266 & 0.1372 & 0.4998 & 0.2322 \\ \midrule 
Piggyback    & 0.0446 & 0.0157 & 0.0236 & 0.5931 & 0.8990 & 0.5100 & 0.0566 & 0.0214 & 0.0302 & 0.0349 & 0.0314 & 0.0322 & 0.2349 & 0.6188 & 0.3129 \\ 
HAT          & 0.0424 & 0.0174 & 0.0279 & 0.5880 & 0.9002 & 0.5126 & 0.0543 & 0.0227 & 0.0372 & 0.0344 & 0.0356 & 0.0317 & 0.2411 & 0.6294 & 0.3296 \\ 
CONURE       & 0.0457 & 0.0169 & 0.0276 & 0.5546 & 0.8967 & 0.5230 & \textbf{0.0598} & 0.0244 & 0.0384 & \textbf{0.0361} & 0.0322 & 0.0305 & 0.2403 & \textbf{0.6391} & 0.3340 \\ \midrule
\proposed    & \textbf{0.0474} & \textbf{0.0189} & \textbf{0.0316} & \textbf{0.6066} & \textbf{0.9048} & \textbf{0.5386} & 0.0577 & \textbf{0.0270} & \textbf{0.0459} & \textbf{0.0361} & \textbf{0.0359} & \textbf{0.0337} & \textbf{0.2444} & 0.6381 & \textbf{0.3354}\\ \bottomrule
\end{tabular}}
\vspace{-1ex}
\label{tab: main_ttl}
\end{table*}

\noindent \textbf{Methods Compared.} 
We mainly compare \proposed~to the most recent state-of-the-art recent method for learning universal user representation via continual learning, i.e., CONURE \cite{yuan2021one}, and various other baseline methods, whose details are given as follows:

\begin{itemize}[leftmargin = 2mm]
    \item \textbf{SinMo} trains a single model for each task  from scratch, and thus no transfer learning occurs between tasks ($M$ models in total).
    
    \item \textbf{FineAll} is pre-trained on task $T_1$ and fine-tuned on each task independently. Note that all the parameters in the backbone network and classifier layers are fine-tuned ($M$ models in total). 
    
    \item \textbf{PeterRec} \cite{yuan2020parameter} is pre-trained on the task $T_1$ and fine-tuned for each task with task-specific model patches and classifier layers. Differently from FineAll, PeterRec needs to maintain only a small number of parameters included in task-specific patches and classifiers, and they are fine-tuned in the subsequent tasks.
    
    \item \textbf{MTL} optimizes the model parameter via multi-task learning. Since not all users in $T_1$ exist in the remaining tasks, we conduct MTL with two objectives, i.e., one for $T_1$ and the other for $T_i (i>1)$.
    
    \item \textbf{Piggyback} \cite{mallya2018piggyback} / \textbf{HAT} \cite{serra2018overcoming} are the continual learning methods proposed in the computer vision domain, which we adapt to the universal user representation.
    \textbf{Piggyback} learns a binary mask for each task after obtaining the model parameters by pre-training on task $T_1$, and \textbf{HAT} learns a soft mask for the output of each layer and allows the entire model parameters to vary during the entire sequence of training tasks.
    
    \item \textbf{CONURE} \cite{yuan2021one} is our main baseline, which learns universal user representations based on the parameter isolation approach.
\end{itemize}

\noindent \textbf{Evaluation Protocol.}
To evaluate the methods, we randomly split each dataset in $T_i$ into train/validation/test data of $80/5/15\%$ following previous work \cite{yuan2021one}. 
We use Mean Reciprocal Rank ($MRR@5$) to measure the model performance on item/query recommendation and product category prediction tasks, and the classification accuracy, i.e., $Acc = \frac{\# \text{Correct predictions}}{\# \text{Total number of users in task}}$, for user profiling tasks. 
We save model parameters for the next task when the performance on validation data gives the best result. Note that the performance of CL-based methods, i.e., Piggyback, HAT, CONURE, and~\proposed, on each task is evaluated after continually training from $T_1$ to $T_M$, i.e., until the end of training sequence.
Furthermore, we measure the effectiveness of capturing the relationship between tasks in terms of forward transfer (FWT) and backward transfer (BWT).
More specifically, we use $FWT^{T_i}=\frac{R^{(T_i,T_i)}-\bar{R}^{T_i}}{\bar{R}^{T_i}}\times 100$ and  $BWT^{T_i}=\frac{R^{(T_M,T_i)}-R^{(T_i,T_i)}}{R^{(T_i,T_i)}}\times 100$, where $R^{(T_j,T_i)}$ is the test performance of the model evaluated on $T_i$ after training on $T_j$, where $j>i$, and $\bar{R}^{T_i}$ is test performance of \textbf{SinMo} on $T_i$.
Note that $FWT^{T_i}>0$ if the performance on $T_i$ is better when it is evaluated after continually training from $T_1$ to $T_i$ compared with the case when it is evaluated after training a single model on $T_i$ from scratch. 
$BWT^{T_i}>0$ if the performance on $T_i$ is better when it is evaluated after continually training from $T_1$ to $T_M$, i.e., until the end of training sequence, compared with the case when it is evaluated after continually training from $T_1$ to $T_i$. 
We provide implementation details in Appendix \ref{app: Implementation Details}.

\begin{table*}[t]
\caption{Model performance on TTL dataset with the original and the reversed task sequence.}
\vspace{-2ex}
\resizebox{0.85 \linewidth}{!}{
\begin{tabular}{c||ccc|ccc|ccc|ccc|ccc|ccc}
\toprule
             \multirow{2}{*}{(a) Original} & \multicolumn{3}{c|}{$T_1$} & \multicolumn{3}{c|}{$T_2$} & \multicolumn{3}{c|}{$T_3$} & \multicolumn{3}{c|}{$T_4$} & \multicolumn{3}{c|}{$T_5$} & \multicolumn{3}{c}{$T_6$} \\ \cline{2-19}
             & MRR@5 & BWT & \multicolumn{1}{c|}{FWT} & MRR@5 & BWT & \multicolumn{1}{c|}{FWT} & MRR@5 & BWT & \multicolumn{1}{c|}{FWT} & ACC & BWT & \multicolumn{1}{c|}{FWT} & ACC & BWT & \multicolumn{1}{c|}{FWT} & ACC & BWT & FWT \\ \midrule \midrule
HAT          & 0.0424 & -11.30\% & - & 0.0174 & -7.45\% & 80.77\% & 0.0279 & -0.71\% & 67.25\% & 0.5880 & -2.52\% & 34.79\% & 0.9002 & -1.98\% & 3.17\% & 0.5126 & - & 17.14\%\\ 
CONURE       & 0.0457 & - & - & 0.0169 & - & 62.50\% & 0.0276 & - & 64.29\% & 0.5546 & - & 23.93\% & 0.8967 & - & 0.74\% & 0.5230 & - & 19.52\%\\ \midrule
\proposed         & \textbf{0.0474} & -0.83\% & - & \textbf{0.0189} & 0.0\% & 81.73\% & \textbf{0.0316} & 3.27\% & 82.13\% & \textbf{0.6066} & 1.23\% & 33.91\% & \textbf{0.9048} & 0.01\% & 1.64\% & \textbf{0.5386} & - & 23.08\%\\ \midrule \midrule
             \multirow{2}{*}{(b) Reversed} & \multicolumn{3}{c|}{$T_1$} & \multicolumn{3}{c|}{$T_6$} & \multicolumn{3}{c|}{$T_5$} & \multicolumn{3}{c|}{$T_4$} & \multicolumn{3}{c|}{$T_3$} & \multicolumn{3}{c}{$T_2$} \\ \cline{2-19}
             & MRR@5 & BWT & \multicolumn{1}{c|}{FWT} & ACC & BWT & \multicolumn{1}{c|}{FWT} & ACC & BWT & \multicolumn{1}{c|}{FWT} & ACC & BWT & \multicolumn{1}{c|}{FWT} & MRR@5 & BWT & \multicolumn{1}{c|}{FWT} & MRR@5 & BWT & FWT \\ \midrule \midrule
HAT          & 0.0422 & -11.72\% & - & 0.5025 & -4.70\% & 20.49\% & 0.8980 & -0.33\% & 1.22\% & 0.5770 & -1.72\% & 31.19\% & 0.0269 & -0.37\% & 60.71\% & 0.0184 & - & 76.92\%\\ 
CONURE       & 0.0457 & - & - & 0.5322 & - & 21.62\% & 0.8849 & - & -0.58\% & 0.5546 & - & 23.93\% & 0.0164 & - & -2.38\% & 0.0119 & - & 14.42\% \\ \midrule 
\proposed         & \textbf{0.0474} & -0.83\% & - & \textbf{0.5365} & 1.84\% & 20.38\% & \textbf{0.9039} & 0.93\% & 0.61\% & \textbf{0.6042} & 0.07\% & 34.92\% & \textbf{0.0313} & 2.62\% & 81.55\% & \textbf{0.0190} & - & 82.69\%\\ \bottomrule
\end{tabular}}
\vspace{-1ex}
\label{tab: task_order}
\end{table*}

\begin{table}[t]
\caption{Model performance and the performance degradation ratio (in bracket) compared to Table \ref{tab: main_ttl} (\%) after training on a noisy task $T'$. }
\vspace{-1ex}
\resizebox{0.9\linewidth}{!}{
\begin{tabular}{c||ccccccc}
\toprule
        & \multicolumn{7}{c}{TTL} \\ \cline{2-8}
        & $T_1$ & $T_2$ & $T_3$ & $T'$ & $T_4$ & $T_5$ & $T_6$ \\ \midrule \midrule
\multirow{2}{*}{HAT} & 0.0411 & 0.0165 & 0.0259 & \multirow{2}{*}{-} & 0.5424 & 0.8870 & 0.4873 \\ 
        & \small{(-3.06 \%)} & \small{(-5.17 \%)} & \small{(-7.16 \%)} &  & \small{(-7.76 \%)} & \small{(-1.47 \%)} & \small{(-4.94 \%)} \\ 
\multirow{2}{*}{CONURE} & 0.0457 & 0.0169 & 0.0276 & \multirow{2}{*}{-} & 0.5245 & 0.8663 & 0.4469 \\ 
        & \small{(0.0 \%)} & \small{(0.0 \%)} & \small{(0.0 \%)} &  & \small{(-5.43 \%)} & \small{(-3.39 \%)} & \small{(-14.55 \%)} \\\midrule
\multirow{2}{*}{\proposed}   & \textbf{0.0472} & \textbf{0.0189} & \textbf{0.0314} & \multirow{2}{*}{-} & \textbf{0.6022} & \textbf{0.9014} & \textbf{0.5312} \\
        & \small{(-0.42 \%)} & \small{(0.0 \%)} & \small{(-0.63 \%)} &  & \small{(-0.73 \%)} & \small{(-0.38 \%)} & \small{(-1.37 \%)} \\\midrule \midrule
& \multicolumn{7}{c}{NAVER Shopping} \\ \cline{2-8}
        & $T_1$ & $T_2$ & $T'$ & $T_3$ & $T_4$ & $T_5$ & $T_6$ \\ \midrule \midrule
\multirow{2}{*}{HAT} & 0.0314 & 0.0302 & \multirow{2}{*}{-} & 0.0309 & 0.2357 & 0.6219 & 0.3180 \\ 
        & \small{(-8.72\%)} & \small{(-15.16\%)} &  & \small{(-2.52\%)} & \small{(-2.24\%)} & \small{(-1.19\%)} &  \small{(-3.51\%)}\\ 
\multirow{2}{*}{CONURE} & \textbf{0.0361} & 0.0322 & \multirow{2}{*}{-} & 0.0291 & 0.2231 & 0.6202 & 0.3122 \\ 
        & \small{(0.0\%)} & \small{(0.0\%)} &  & \small{(-4.59\%)} & \small{(-7.16\%)} & \small{(-2.95\%)} & \small{(-6.53\%)} \\ \midrule
\multirow{2}{*}{\proposed}  & 0.0346 & \textbf{0.0336} & \multirow{2}{*}{-} & \textbf{0.0329} & \textbf{0.2378} & \textbf{0.6348} & \textbf{0.3329} \\ 
        & \small{(-4.15\%)} & \small{(-6.41\%)} &  & \small{(-2.37\%)} & \small{(-2.7\%)} & \small{(-0.52\%)} & \small{(-0.75\%)} \\ \bottomrule
\end{tabular}}
\vspace{-3ex}
\label{tab: noisy task}
\end{table}

\vspace{-1ex}
\subsection{Overall Performance}
The experimental results of sequential learning on the various tasks in three datasets are given in Table \ref{tab: main_ttl}.
We have the following observations:
\textbf{1)} Positive transfer between tasks exists when comparing SinMo to other baseline methods (See also Table~\ref{tab: task_order} (a)).
Specifically, TL-based approaches, i.e., FineAll and PeterRec, outperform SinMo in most of the cases, indicating that the knowledge obtained from task $T_1$ is helpful in learning task $T_i~(i > 1)$.
Moreover, CL-based methods, i.e., HAT, CONURE, and \proposed, outperform the TL-based methods verifying that the knowledge transfer happens not only between paired tasks but also between multiple tasks.
On the other hand, positive transfer rarely happens when multiple tasks are simultaneously trained (See SinMo vs. MTL).
These results indicate that the online web platform, which contains multiple tasks in nature, requires continual user representations for the true understanding of users.
\textbf{2)} \proposed~outperforms the CL-based approaches that do not consider the relationship between the tasks, i.e., Piggyback, HAT, and CONURE.
This is because learning the relationship between the tasks, i.e., whether the tasks are positively related or not, can further accelerate the knowledge transfer between the tasks, which has been shown to exist in our first observation above.
\textbf{3)} Moreover, we observe a positive backward transfer occurs when training \proposed~in Table \ref{tab: task_order} (a), which can never happen in parameter isolation-based approaches, i.e., CONURE (hence, BWT of CONURE is not reported).
This is because our model allows the entire model parameters to be modified during the entire training sequence, enabling the knowledge obtained from the new tasks to be transferred to the previous tasks.
\textbf{4)} However, allowing the model parameters to be modified also incurs a severe performance degradation as a new task arrives, i.e., catastrophic forgetting, if there exists no module specifically designed to alleviate the issue (see HAT in Table \ref{tab: task_order} (a)).
This also verifies that \proposed~successfully alleviates catastrophic forgetting with the knowledge retention module, which facilitates the model to retain the knowledge obtained from the previous tasks.

\subsection{Comparison to Parameter Isolation}
In this section, we verify the effectiveness of \proposed~compared to the parameter isolation-based approach, i.e., CONURE, by conducting experiments on the different sequences of tasks in Table \ref{tab: task_order}, and inserting noisy tasks among the sequence of tasks in Table \ref{tab: noisy task}.

\smallskip
\noindent\textbf{\proposed~is robust to the change of task orders.}
As shown in Table \ref{tab: task_order} (b), our model maintains its performance even when the task sequence is trained in the reversed direction, i.e., $T_1 \rightarrow T_6 \rightarrow T_5 \rightarrow \cdots \rightarrow T_2$, while the CONURE's performance in task $T_2$ and $T_3$ severely deteriorates.
As we mentioned in Sec.~\ref{Sec: Parameter Isolation}, this is because the number of trainable parameters in CONURE decreases as the number of trained tasks increases.
Therefore, when a large portion of model parameters are trained for learning less informative tasks in the early stage, the model cannot learn the subsequent tasks that actually require a large number of parameters for generalization.
On the other hand, \proposed~successfully learns the subsequent tasks by maintaining its learning capacity during the entire sequence of training.
Considering that the sequence of tasks cannot be arbitrarily determined in the real world, we argue that \proposed~is more practical than CONURE in real-world applications.

\noindent\textbf{\proposed~is robust to the negative transfer.}
\looseness=-1
To investigate the impact of negative transfer between the tasks, we conduct experiments by inserting an uninformative task among the sequence of tasks.
Specifically, given the total user set $\mathcal{U}$, we randomly sample 50\% of users and generate a random label of 50 classes for each user to create a noisy task $T'$.
In Table \ref{tab: noisy task}, we have the following observations:
\textbf{1)} Compared to Table \ref{tab: main_ttl}, the overall performance of all methods degrade due to the effect of noisy task $T'$.
This indicates that a negative transfer occurs between tasks that have no relationship between one another.
However, thanks to the parameter isolation approach, CONURE does not experience a performance drop in the tasks trained before the noisy task.
\textbf{2)} On the other hand, we observe that CONURE suffers from the largest performance drop among the methods in the tasks that are trained after the noisy task.
This is because the model parameters that are disrupted while training the noisy task are frozen after the task, impacting the performance of the subsequent tasks.
\textbf{3)} Moreover, we observe that \proposed~successfully alleviates the effect of the noisy task with a moderate performance drop, which is in contrast to HAT.
We attribute this to our proposed relation-aware task-specific masks, which automatically disregard the information from the noisy task making \proposed~robust to the negative transfer.

To summarize our findings, \proposed~is robust to the change of task orders and the negative transfer from noisy tasks, demonstrating the importance of 1) maintaining the learning capacity and 2) the relation-aware task-specific masks.
Moreover, considering that there exist multiple tasks of unknown sequence or negative correlation in real-world web platforms, we argue that \proposed~is practical in reality.
We further analyze the universal user representation obtained by the methods Figure \ref{fig: finetune} in Appendix \ref{app: Evaluating Universal User Representation}.

\subsection{Model Analysis}
\noindent \textbf{Effect of Relation-aware Task-specific Masks.}
We investigate the importance of the relation-aware task-specific masks by comparing various masking strategies.
Specifically, we compare the proposed relation-aware task-specific masks defined in Eq.~\ref{eq:relation_aware} with the masks that do not aggregate the information from the other tasks, i.e., \textit{w/o relation}, defined in Eq.~\ref{eq:mask_ours}, and the mask without the opposite relatedness, i.e., \textit{only positive} (no tanh$(-s\cdot\mathbf{e}_{k}^{T})$ in Eq.~\ref{eq:p_kT}).
Based on Table \ref{tab: task embedding}, we have the following observations:
\textbf{1)} The masks that do not have the task-relation information between the tasks perform the worst (i.e., \textit{w/o relation}), indicating that injecting the relationship between tasks into the task-specific masks is helpful for transferring knowledge between the tasks.
\textbf{2)} On the other hand, by comparing the BWT ratio between \textit{only positive} and \proposed, we observe that modeling the opposite relatedness between the tasks (see Eq.~\ref{eq:p_kT}) is crucial for alleviating catastrophic forgetting.
This is because the opposite relatedness provides further relational information between the tasks, making it easy for the model to decide which dimension of the task embedding to amplify or suppress.
In summary, our proposed relation-aware task-specific masks not only successfully transfer the knowledge between the tasks but also retain the previously obtained knowledge. Refer to Figure \ref{fig: masking} in Appendix \ref{app: Masking Layer Output} for other masking strategies.

\smallskip
\noindent \textbf{Effect of Relation-aware User Sampling.}
\looseness=-1
We verify the effectiveness of the user sampling in terms of the model performance and training time in Table \ref{tab: sampling ratio}.
To do so, we compare the performance of the automatic sampling strategy described in Eq.~\ref{eq: rho} with a variant user sampling strategy, i.e., $\rho_{i,j}=\rho_{min}$ that selects the minimum value among $\rho_{i,j}$ for all $i$ and $j$ when training task $T_i$.
Moreover, we compare it to the one that generates pseudo-labels for all the users in the task, i.e., no sampling, which is equivalent to setting $\rho_{i,j} = 1.0$.
In Table \ref{tab: sampling ratio}, we observe that $\rho_{i,j}=1.0$ generally performs the best, which is expected as all the users in every task is used to train the model. However, we observe that our proposed sampling strategy in Eq.~\ref{eq: rho} shows a competitive performance even with a small number of users in each task and reduced training time, which shows the efficiency of our proposed relation-aware user sampling strategy\footnote{As mentioned in Sec.~\ref{Sec: Relation-aware User Sampling}, the value of $\rho_{i,j}$ is between 0.0007 and 0.08, which means only 0.7\% to 8\% of the users are required to achieve the performance shown in Table \ref{tab: sampling ratio}.}. Finally,  $\rho_{i,j}=\rho_{min}$ performs the worst, which implies that considering the relationship between the tasks is beneficial when sampling users.
Refer to Table \ref{app tab: sampling ratio ecom}, \ref{tab: sampling reversed} in Appendix \ref{app: Effect of Relation-aware User Sampling} for additional experiments on user sampling.

\begin{table}[t]
\caption{Model performance and BWT ratio (in bracket) with various masking strategies on TTL dataset.}
\vspace{-2ex}
\resizebox{0.9\linewidth}{!}{
\begin{tabular}{c||cccccc}
\toprule
         & $T_1$ & $T_2$ & $T_3$ & $T_4$ & $T_5$ & $T_6$ \\ \midrule\midrule
\textit{w/o}  &  0.0455  &  0.0182  &  0.0292  &  0.5940  &  0.8999  &  0.5354  \\
\textit{relation} &  \small{(-4.81\%)} & \small{(-1.09\%)} & \small{(-1.02\%)} & \small{(-0.26\%)} & \small{(-0.06\%)}  &\small{(0.0\%)}\\ \midrule
\textit{only} &  0.0471  &  0.0185  &  0.0293  &  0.6054  &  0.9023  &  0.5289  \\ 
\textit{positive} & \small{(-1.46\%)}  & \small{(-1.07\%)}  &  \small{(0.69\%)} & \small{(-1.11\%)}  & \small{(0.21\%)} &\small{(0.0\%)}\\ \midrule
\multirow{2}{*}{\proposed}  & \textbf{0.0474} &  \textbf{0.0189} & \textbf{0.0316} &  \textbf{0.6066}  &  \textbf{0.9048}  &  \textbf{0.5386}  \\
                        & \small{(-0.83\%)}  &  \small{(0.0\%)} & \small{(3.27\%)} & \small{(1.23\%)}  &  \small{(0.01\%)} & \small{(0.0\%)}\\ \midrule
\end{tabular}}
\vspace{-2ex}
\label{tab: task embedding}
\end{table}

\begin{table}[t]
\caption{Model performance and training time (i.e., sec/epoch) (in bracket) over various sampling strategies on TTL dataset.}
\vspace{-2ex}
\resizebox{0.9\linewidth}{!}{
\begin{tabular}{c|c||cccccc}
\toprule
    & \small{Sampling}& $T_1$     & $T_2$     & $T_3$     & $T_4$     & $T_5$     & $T_6$     \\ \midrule\midrule
\multirow{2}{*}{$\rho_{i,j}=\rho_{min}$}&\multirow{2}{*}{\cmark} & 0.0470 & 0.0184 & 0.0280 & 0.6027 & 0.9007 & 0.5385 \\ 
    & & \small{( - )} & \small{(625.47)} & \small{(77.82)} & \small{(417.65)} & \small{(510.80)} & \small{(414.44)} \\ \midrule 
   \multirow{2}{*}{$\rho_{i,j}=$ Eq.\ref{Eq:sampling}} &\multirow{2}{*}{\cmark}& 0.0474 & 0.0189 & \textbf{0.0316} & 0.6066 & \textbf{0.9048} & 0.5386 \\ 
    & & \small{( - )} & \small{(625.47)} & \small{(90.79)} & \small{(504.3)} & \small{(583.77)} & \small{(494.14)} \\ \midrule 
\multirow{2}{*}{$\rho_{i,j}=1.0$} &\multirow{2}{*}{\xmark} & \textbf{0.0475} & \textbf{0.0190} & 0.0313 & \textbf{0.6143} & 0.9047 & \textbf{0.5403} \\ 
 & & \small{( - )} & \small{(1146.70)} & \small{(151.32)} & \small{(1179.31)} & \small{(1355.18)} & \small{(797.09)} \\  \bottomrule
\end{tabular}}
\label{tab: sampling ratio}
\vspace{-2ex}
\end{table}

\section{Conclusion}
In this paper, we propose a novel continual user representation learning model, named \proposed.
The main idea is to utilize task embeddings for generating relation-aware task-specific masks that enable the model to maintain the learning capability during training and facilitate the relationship between the tasks to be captured.
By doing so, \proposed~successfully transfers the knowledge between the tasks not only in the forward direction but also backward direction. Moreover, \proposed~ prevents catastrophic forgetting through a novel knowledge retention module.
Our extensive experiments on various real-world datasets demonstrate that \proposed~is not only effective and efficient, but also robust to the change of task orders and negative transfer, which shows the practicality of~\proposed~ in real-world applications.

\smallskip \noindent \textbf{Acknowledgements} This work was supported by NAVER Corporation, and Institute of Information \& communications Technology Planning \& Evaluation (IITP) grant funded by the Korea government(MSIT) (No.2022-0-00157).


\clearpage

\bibliographystyle{ACM-Reference-Format}
\balance
\bibliography{sample-base}

\clearpage
\appendix

\section{NAVER Shopping Dataset}
\label{app: A Proprietary Dataset}

In this paper, we use a proprietary dataset that is collected at an e-commerce platform, i.e., NAVER Shopping, from 07/01/2022 to 11/30/2022. 
We randomly selected a subset of users from that period whose search query history is longer than 10, 
and the queries that are searched more than 200 times. 
That is, user search behavior sequences used in this paper have queries searched over 200 times and a user sequence length greater than 10.
The e-commerce platform contains a multitude of services that are not limited to online shopping, e.g., search engines and news.
We process the datasets to create a sequence of tasks, regarding the interest and needs of the platform.
NAVER Shopping consists of two search query prediction tasks in the portal whose queries are not limited to the shopping domain,
two purchased item category prediction tasks in the shopping platform, and two tasks for predicting users' gender and age. 
We have detailed statistics and descriptions in Table \ref{tab: Ecom Data Statistics} and below:

\begin{table}[h]
\caption{Statistics for NAVER Shopping dataset.}
\resizebox{1.0\linewidth}{!}{
\begin{tabular}{wc{1cm}|wc{4cm}wc{1.3cm}wc{1.3cm}wc{3cm}}
    \toprule
          & \multirow{2}{*}{Task} & \multirow{2}{*}{\# Users} & \# Unique & \multirow{2}{*}{Date} \\
          &  &  & labels &  \\ \midrule \midrule
    $T_1$ & Next Search Query & 0.9M & 0.58M & $07/01/2022 \sim 10/31/2022$ \\
    $T_2$ & Next Search Query & 0.59M & 0.51M & $11/01/2022 \sim 11/30/2022$ \\
    $T_3$ & Minor Item Category & 0.15M & 4K & $11/01/2022 \sim 11/30/2022$ \\
    $T_4$ & Major Item Category & 0.15M & 10 & $11/01/2022 \sim 11/30/2022$ \\
    $T_5$ & Gender & 0.82M & 2 & - \\
    $T_6$ & Age & 0.82M & 9 & - \\ \bottomrule
\end{tabular}}
\label{tab: Ecom Data Statistics}
\end{table}

\begin{itemize}[leftmargin = 5mm]
    \item \textbf{$T_1$} consists of the users' recent search queries in the online portal platform during the period of 07/01/2022 to 10/31/2022. 
    We use a user's 60 recent search queries, and train the model to predict the next search query in an autoregressive manner.
    \item \textbf{$T_2$} consists of the next five search queries of the search queries contained in $T_1$. 
    That is, search queries in $T_2$ are collected during the period of 11/01/2022 to 11/30/2022.
    Given the recent 60 search queries of the user in $T_1$, the model is trained to predict the user's next five search queries.
    \item \textbf{$T_3$} and \textbf{$T_4$} are item category prediction tasks with a different hierarchy.
    Note that the items in NAVER Shopping dataset are hierarchically categorized, i.e., \textbf{major} categories $\rightarrow$ \textbf{middle} categories $\rightarrow$ \textbf{small} categories $\rightarrow$ \textbf{minor} categories.
    Therefore, given a user's 60 recent search queries, the model is trained to predict \textbf{minor} and \textbf{major} categories of the items during the period of 2022/11/01 to 2022/11/30 in $T_3$ and $T_4$, respectively.
    It is also worth noting that $T_3$ and $T_4$ can be recognized as cross-domain recommendations, which is a common interest in industry,  since the model aims to recommend \textbf{shopping items} based on the user's \textbf{search query history}.
    \item \textbf{$T_5$} is a gender prediction task based on the user's recent 60 search queries.
    \item \textbf{$T_6$} is an age prediction task based on the user's recent 60 search queries.
\end{itemize}

\section{Additional Experiments}

\subsection{Effect of Relation-aware User Sampling}
\label{app: Effect of Relation-aware User Sampling}

In Table~\ref{app tab: sampling ratio ecom}, we additionally conduct experiments on the effect of relation-aware user sampling in NAVER Shopping dataset.
We observe that \proposed~efficiently learns from a portion of the users with a moderate performance degradation.

\begin{table}[h]
\caption{Model performance and training time (i.e., sec/epoch) (in bracket) over various sampling ratios on NAVER Shopping dataset.}
\resizebox{1.0\linewidth}{!}{
\begin{tabular}{c||cccccc}
\toprule
    & $T_1$     & $T_2$     & $T_3$     & $T_4$     & $T_5$     & $T_6$     \\ \midrule\midrule
\multirow{2}{*}{$\rho_{min}$} & 0.0358 & 0.0350 & 0.0323 & 0.2433 & 0.6379 & 0.3352 \\ 
    & \small{(-)} & \small{(417.71)} & \small{(111.60)} & \small{(62.06)} & \small{(215.25)} & \small{(248.26)} \\ \midrule
\textit{w/o} & \textbf{0.0363} & 0.0361 & 0.0342 & 0.2485 & 0.6401 & 0.3352 \\ 
\textit{sampling} & \small{(-)} & \small{(626.57)} & \small{(172.41)} & \small{(133.45)} & \small{(428.74)} & \small{(446.47)} \\ \midrule
\multirow{2}{*}{\proposed} & 0.0361 & 0.0359 & 0.0337 & 0.2444 & 0.6381 & 0.3354 \\ 
    & \small{(-)} & \small{(417.71)} & \small{(135.13)} & \small{(79.56)} & \small{(242.16)} & \small{(278.95)} \\ \bottomrule
\end{tabular}}
\label{app tab: sampling ratio ecom}
\end{table}

\noindent Moreover, we also investigate the effect of relation-aware user sampling in the TTL dataset of reverse order in Table \ref{tab: sampling reversed}.
We observe that relation-aware user sampling approaches also maintain the model's robustness to the task orders.

\begin{table}[h]
\caption{Model performance and training time (i.e., sec/epoch) (in bracket) over various sampling ratios on TTL dataset Reversed task sequence.}
\resizebox{1.0\linewidth}{!}{
\begin{tabular}{c|c||cccccc}
\toprule
    & \small{Sampling}& $T_1$     & $T_6$     & $T_5$     & $T_4$     & $T_3$     & $T_2$     \\ \midrule\midrule
\multirow{2}{*}{$\rho_{i,j}=\rho_{min}$} &\multirow{2}{*}{\cmark}& 0.0471 & 0.5304 & 0.9010 & 0.6011 & 0.0307 & 0.0188 \\ 
    & & \small{( - )} & \small{(393.53)} & \small{(491.11)} & \small{(449.84)} & \small{(92.44)} & \small{(743.94)} \\ \midrule 

\multirow{2}{*}{$\rho_{i,j}=$ Eq.\ref{Eq:sampling}} &\multirow{2}{*}{\cmark}& 0.0474 & 0.5365 & \textbf{0.9039} & 0.6042 & \textbf{0.0313} & 0.0190 \\ 
    && \small{( - )} & \small{(393.53)} & \small{(548.92)} & \small{(517.22)} & \small{(108.98)} & \small{(873.98)} \\ \midrule
    \multirow{2}{*}{$\rho_{i,j}=1.0$} & \multirow{2}{*}{\xmark} &\textbf{0.0475} & \textbf{0.5366} & 0.9031 & \textbf{0.6104} & 0.0311 & \textbf{0.0192} \\ 
 & & \small{( - )} & \small{(568.11)} & \small{(989.42)} & \small{(1133.54)} & \small{(197.68)} & \small{(1673.55)} \\ \midrule \bottomrule
\end{tabular}}
\label{tab: sampling reversed}
\end{table}

\subsection{Evaluating Universal User Representation}
\label{app: Evaluating Universal User Representation}

\begin{table*}[!]
\caption{Hyperparameter specifications of \proposed}
\resizebox{0.8\linewidth}{!}{
\begin{tabular}{c||cccccc||ccc||cccccc}
\toprule
             &\multicolumn{6}{c||}{TTL} & \multicolumn{3}{c||}{ML} &\multicolumn{6}{c}{NAVER Shopping}\\ \cline{2-16}
             & $T_1$ & $T_2$ & $T_3$ & $T_4$ & $T_5$ & $T_6$ & $T_1$ & $T_2$ & $T_3$ & $T_1$ & $T_2$ & $T_3$ & $T_4$ & $T_5$ & $T_6$\\ \midrule\midrule
 Learning& \multirow{2}{*}{0.01} & \multirow{2}{*}{0.0001} & \multirow{2}{*}{0.0001} & \multirow{2}{*}{0.0001} & \multirow{2}{*}{0.0001} & \multirow{2}{*}{0.0001} & \multirow{2}{*}{0.001} & \multirow{2}{*}{0.0001} & \multirow{2}{*}{0.0005} & \multirow{2}{*}{0.001} & \multirow{2}{*}{0.0001} & \multirow{2}{*}{0.0001} & \multirow{2}{*}{0.0005} & \multirow{2}{*}{0.0005} & \multirow{2}{*}{0.0005} \\
rate ($\eta$) & & & & & & & & & & & & & & & \\\midrule
Scaling & \multirow{2}{*}{-} & \multirow{2}{*}{6.0} & \multirow{2}{*}{6.0} & \multirow{2}{*}{6.0} & \multirow{2}{*}{6.0} & \multirow{2}{*}{6.0} & \multirow{2}{*}{-} & \multirow{2}{*}{6.0} &  \multirow{2}{*}{6.0} & \multirow{2}{*}{-} & \multirow{2}{*}{6.0} & \multirow{2}{*}{6.0} & \multirow{2}{*}{6.0} & \multirow{2}{*}{6.0} & \multirow{2}{*}{6.0} \\
($c$) & & & & & & & & & & & & & & & \\ \midrule
Positive& \multirow{2}{*}{50} & \multirow{2}{*}{50} & \multirow{2}{*}{50} & \multirow{2}{*}{50} & \multirow{2}{*}{50} & \multirow{2}{*}{50} & \multirow{2}{*}{50} & \multirow{2}{*}{50} & \multirow{2}{*}{50} & \multirow{2}{*}{50} & \multirow{2}{*}{50} & \multirow{2}{*}{50} & \multirow{2}{*}{50} & \multirow{2}{*}{50} & \multirow{2}{*}{50} \\
scaling ($s_\text{max}$)& & & & & & & & & & & & & & &\\\midrule
Knowledge & \multirow{2}{*}{-}      & \multirow{2}{*}{0.7} & \multirow{2}{*}{0.7} & \multirow{2}{*}{0.7} & \multirow{2}{*}{0.7} & \multirow{2}{*}{0.7} & \multirow{2}{*}{-}      & \multirow{2}{*}{0.7} &  \multirow{2}{*}{0.9} & \multirow{2}{*}{-} & \multirow{2}{*}{0.9} & \multirow{2}{*}{0.9} & \multirow{2}{*}{0.7} & \multirow{2}{*}{0.7} & \multirow{2}{*}{0.9} \\
retention ($\alpha$) & & & & & & & & & & & & & & &\\\midrule 
Batch size & \multirow{2}{*}{32} & \multirow{2}{*}{1024} & \multirow{2}{*}{1024} & \multirow{2}{*}{1024} & \multirow{2}{*}{1024} & \multirow{2}{*}{1024} & \multirow{2}{*}{32} & \multirow{2}{*}{1024} & \multirow{2}{*}{1024} & \multirow{2}{*}{64} & \multirow{2}{*}{1024} & \multirow{2}{*}{1024} & \multirow{2}{*}{1024} & \multirow{2}{*}{1024} & \multirow{2}{*}{1024} \\ 
($b$) & & & & & & & & & & & & & & &\\
\bottomrule
\end{tabular}}
\label{tab: hyperparameter}
\end{table*}

\begin{figure}[h]
    \centering
    \includegraphics[width=0.9\linewidth]{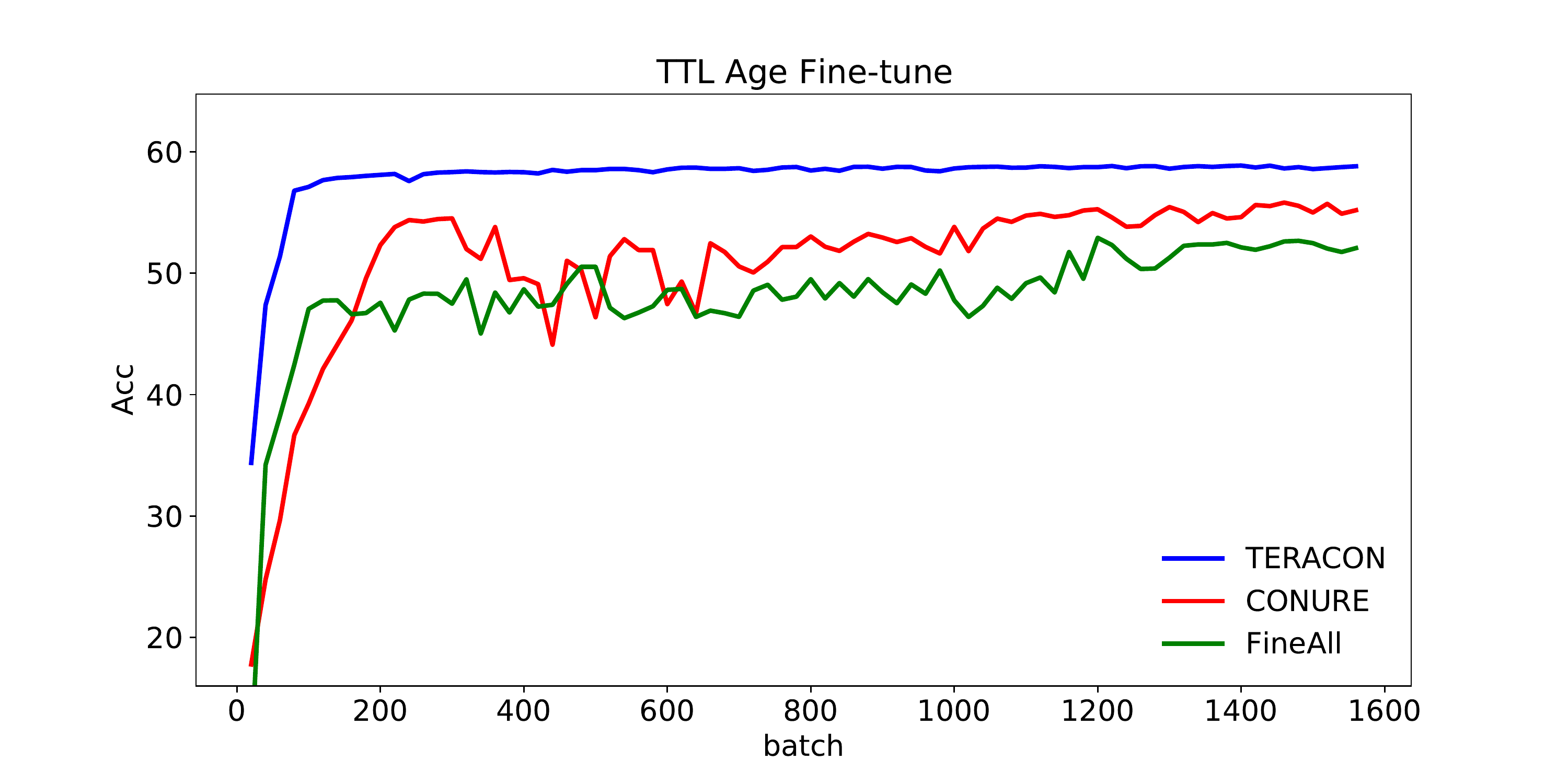}
    \caption{Model performance per epoch on the age prediction task.}
    \label{fig: finetune}
\end{figure}

In this section, we empirically evaluate the quality of universal user representations by comparing how the representation adapts to the new tasks in Figure \ref{fig: finetune}.
More specifically, we train the models on the sequence of tasks $T_1, T_2$, and $T_3$ in TTL dataset, and then evaluate how the embeddings obtained from each model adapt to the task $T_4$, i.e., age prediction task.
We have the following observations:
\textbf{1)} By comparing the initial performance between the methods, we argue that the universal user representation obtained by \proposed~provides better initial points for the new tasks compared to the baseline methods.
\textbf{2)} Moreover, thanks to the good initial points, our model converges to the optimal point much faster than the baseline methods.
Therefore, we argue that \proposed~truly learns the universal user representation that can be easily adapted to the various tasks, which will bring further practicality to web platform applications.

\subsection{Advantages of Masking Layer Output}
\label{app: Masking Layer Output}

\begin{figure}[h]
    \centering
    \includegraphics[width=0.9\linewidth]{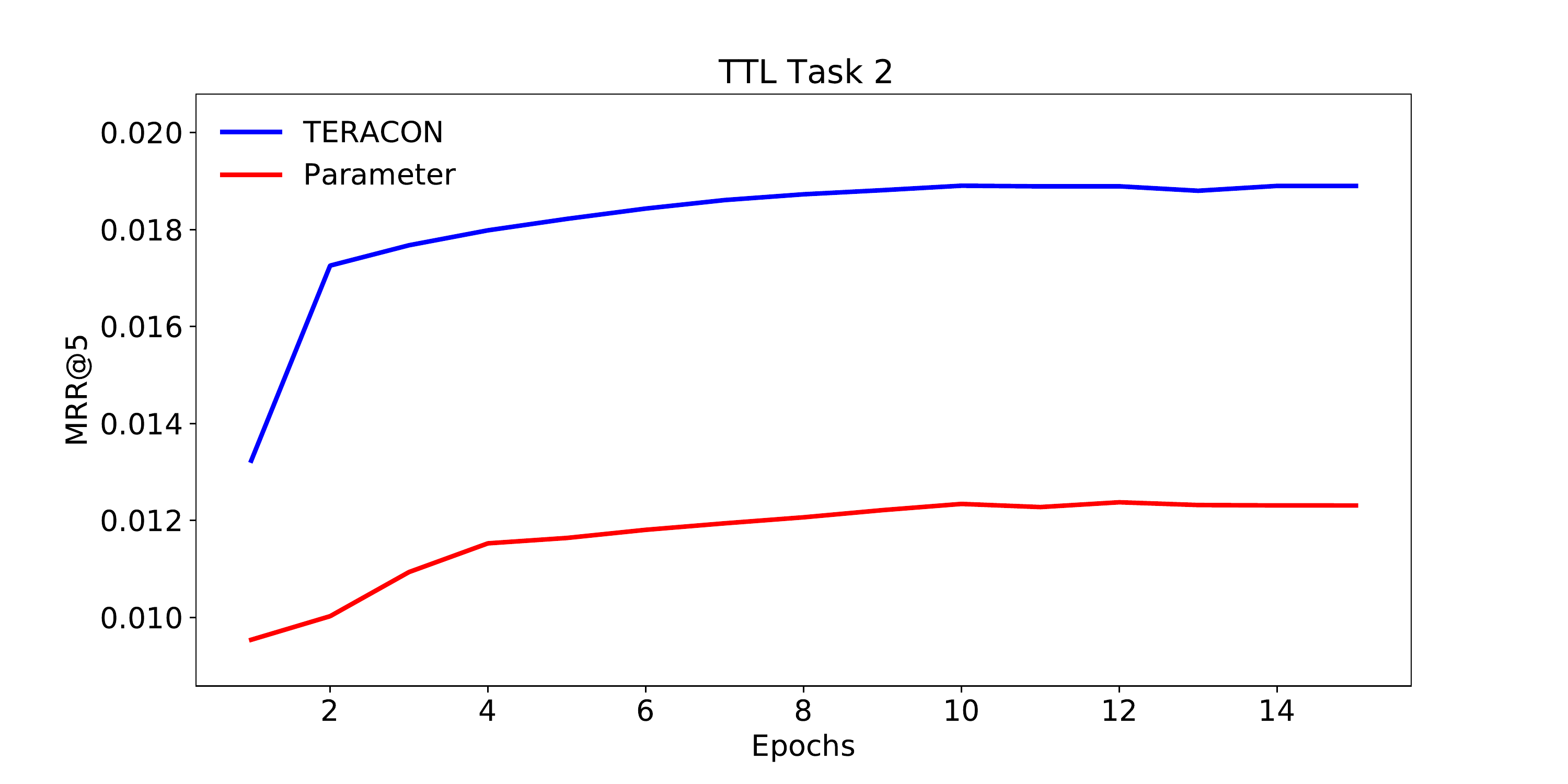}
    \caption{Comparison on the parameter masking and \proposed.}
    \label{fig: masking}
\end{figure}

In this section, we investigate the advantages of masking the outputs of each layer compared to directly masking the model parameters by measuring the test performance during the whole training epochs in Figure \ref{fig: masking}.
We observe that model that masks the whole parameters, i.e., Parameter, has a worse initial point and converges slowly compared to \proposed.
This is because masking all parameters require the same number of parameters for masking since the masking operation is done via an elementwise product operation, which will make it difficult to train the model.
On the other hand, \proposed~requires a small portion of parameters for masking, which is beneficial in model training and computational cost.

\section{Implementation Details}
\label{app: Implementation Details}

For a fair comparison, we set the dimension of item and task embeddings to 256 for all the methods and datasets.
All the tasks are conducted 1024 batch size ($b$) but for $T_1$, due to the limitations of GPU memory, we use a smaller batch size ($b$).
Moreover, we use the Adam optimizer to train the models in all tasks.
For hyperparameters, we tune the model in certain ranges as follows: learning rate $\eta$ in $\left\{0.001, 0.0005, 0.0001, 0.00005\right\}$, scaling $c$ in $\left\{3.0, 4.5, 6.0\right\}$, positive scaling $s_\text{max}$ in $\left\{5, 50, 100, 500\right\}$, knowledge retention coefficient $\alpha$ in $\left\{0.1, 0.3, 0.5, 0.7, 0.9, 1.0\right\}$. 
We report the best performing hyperparameters for all the tasks in each dataset in Table \ref{tab: hyperparameter}.

For the baseline methods, we tried our best to reproduce the results in their own papers by following their desciption on implementation details.
Specifically, we follow the pruning ratios reported on CONURE for each task, i.e., $70/80/90/80/90/90\%$ and $70/80/70\%$, for TTL and ML, respectively.
Moreover, for NAVER Shopping dataset, we tune the pruning ratio between $70 \sim 90 \%$ for each task.

We use NVIDIA GeForce A6000 48GB for TTL and ML dataset, and use eight Tesla P40 for NAVER Shopping datasets.

\subsection{Reproducibility}
\label{App: Reproducibility}
For baseline models, we use the official codes published by authors as shown in Table \ref{tab: codes}, and then conduct evaluations within the same environment. Refer to \textbf{\textit{Souce code link}} for our source code and instructions on how to run the code to reproduce the results reported in the experiments.

\begin{table}[h]
    \small
    \centering
    \caption{Source code links of the baseline methods.}
    \begin{tabular}{c|c}
    Methods & Source code \\ \hline \hline
    CONURE & \url{https://github.com/yuangh-x/2022-NIPS-Tenrec} \\
    PeterRec & \url{https://github.com/yuangh-x/2022-NIPS-Tenrec} \\
    HAT & \url{https://github.com/joansj/hat} \\
    Piggyback & \url{https://github.com/arunmallya/piggyback} \\
    NextitNet & \url{https://github.com/syiswell/NextItNet-Pytorch}\\
    \hline
    \proposed & \url{https://github.com/Sein-Kim/TERACON} \\
    \hline
    \end{tabular}
    \label{tab: codes}
\end{table}

\end{document}